\documentclass[submission, Phys]{SciPost}

\usepackage{amsfonts,amssymb,amsmath,enumerate}
\usepackage{color,ulem}

\usepackage{xcolor,cancel}

\newtheorem{thm}{Theorem}[section]
\newtheorem{prop}[thm]{Proposition}
\newtheorem{lemma}[thm]{Lemma}
\newtheorem{coro}[thm]{Corollary}
\newtheorem{defi}[thm]{Definition}
\newcommand{\qed}{{\hfill$\square$\\}}

\newtheorem{rmk}{Remark}[section]

\newcommand{\bea}{\begin{eqnarray}}
\newcommand{\eea}{\end{eqnarray}}
\newcommand{\beano}{\begin{eqnarray*}}
\newcommand{\eeano}{\end{eqnarray*}}
\newcommand{\beq}{\begin{equation}}
\newcommand{\eeq}{\end{equation}}

\newcommand{\mb}[1]{\hspace{2.1ex}\mbox{#1}\hspace{2.1ex}}
\numberwithin{equation}{section}

\newcommand\atopn[2]{\genfrac{}{}{0pt}{}{#1}{#2}}

 \def\ff{{\mathfrak f}}       \def\fl{{\mathfrak l}}
\def\fg{{\mathfrak g}}

            \def\fo{{\mathfrak o}}

    \def\cB{{\cal B}}

        \def\cO{{\cal O}}
\def\cP{{\cal P}}

\newcommand{\BB}{{\mathbb B}}
\newcommand{\CC}{{\mathbb C}}

\newcommand{\FF}{{\mathbb F}}
\newcommand{\II}{{\mathbb I}}

    \def\sT{\mathsf{T}}

\def\sh{\mathsf{t}}

\def\proof{\textit{Proof:\ }}

\newcommand{\so}{\scriptscriptstyle \rm I}
\newcommand{\st}{\scriptscriptstyle \rm I\hspace{-1pt}I}
\newcommand{\sth}{\scriptscriptstyle \rm I\hspace{-1pt}I\hspace{-1pt}I}

\def\ee{\mathsf{e}}
\def\bu{{\bar u}}
\def\bv{{\bar v}}
\def\bw{{\bar w}}
\def\bx{{\bar x}}

\def\su#1{\bu^{(#1)}}
\def\sv#1{\bv^{(#1)}}
\def\sw#1{\bw^{(#1)}}
\def\sx#1{\bx^{(#1)}}

\def\r#1{(\ref{#1})}

\def\sk#1{\left(#1\right)}
\def\rvec{|0\rangle}
\def\lvec{\langle0|}
\def\t{p}

\begin{document}

\setcounter{page}{0}

\vspace{12pt}

\begin{center}
\begin{Large}
{\bf Scalar products and norm of Bethe vectors \\[1ex]
in $\mathfrak{o}_{2n+1}$ invariant integrable models }
\end{Large}

\vspace{20pt}

\begin{large}
A.~Liashyk${}^{a}$, S.~Pakuliak${}^{b}$ and E.~Ragoucy${}^{b}$
\end{large}

\vspace{10mm}

${}^a$ {\it Beijing Institute of Mathematical Sciences and Applications (BIMSA),\\
No. 544, Hefangkou Village Huaibei Town, Huairou District Beijing 101408, China}

\vspace{2mm}

${}^b$ {\it Laboratoire d'Annecy-le-Vieux de Physique Théorique (LAPTh)\\
Chemin de Bellevue, BP 110, F-74941, Annecy-le-Vieux cedex, France}
\vspace{2mm}

E-mails: liashyk@bimsa.cn, pakuliak@lapth.cnrs.fr, ragoucy@lapth.cnrs.fr

\end{center}

\vspace{5mm}

\begin{center}
\begin{minipage}{14cm}
\centerline{\bf Abstract}\smallskip
{\footnotesize
We compute scalar products of off-shell Bethe vectors in 
models with $\fo_{2n+1}$ symmetry. The scalar products are expressed as a 
sum over partitions of the Bethe parameter sets, the building blocks being the 
so-called highest coefficients.
We prove some recurrence relations and a residue theorem for these 
highest coefficients, and prove that they are consistent with the 
reduction to $\fg\fl_n$ invariant models.
We also express the norm of on-shell Bethe vectors as a Gaudin determinant.
}
\end{minipage}
\end{center}

\setcounter{footnote}{0}

\medskip

\section{Introduction}
Integrable systems, characterized by their infinite number of conserved 
quantities, occupy a central place in mathematical physics. These systems, 
appearing prominently in models of statistical mechanics and quantum field 
theory, exhibit remarkable solvability due to their underlying algebraic 
structures.
For instance, in quantum many-body physics, the Lieb–Liniger \cite{LiebL63, Lieb63} model describes a one-dimensional Bose gas with delta-function interactions. It is exactly solvable and has become a cornerstone in the study of cold atomic gases, particularly in regimes where quantum fluctuations dominate. Experiments with quasi-1D Bose–Einstein condensates have directly verified the predictions of this model, including properties like the Tonks–Girardeau \cite{Tonks36, Gir60} gas behavior.
In quantum field theory, integrable systems appear in two-dimensional models, such as the sine-Gordon and Thirring models \cite{Thir58}. The sine-Gordon model, for instance, supports solitonic excitations with an exact S-matrix \cite{ZZ78}. This has implications for understanding non-perturbative effects and dualities.
In statistical mechanics, the 2D Ising model \cite{Ons44} is an exactly solvable lattice model that provides exact results for phase transitions. 
In string theory and AdS/CFT correspondence, integrability plays a vital role. The planar limit of $N=4$ super Yang–Mills theory is conjectured to be integrable, allowing for the computation of anomalous dimensions of operators via integrable spin chain techniques  \cite{MZ03}. This has led to the development of the so-called integrability bootstrap program for solving gauge theories in the planar limit.
A key tool for solving these systems is the Bethe Ansatz, first 
introduced by Hans Bethe in 1931 in the context of the Heisenberg spin 
chain, which provides exact solutions for eigenstates of certain 
Hamiltonians \cite{Bethe}.

The Bethe Ansatz leads to the construction of Bethe vectors 
(see \cite{KR} for $\mathfrak{gl}_N$ and \cite{Res91} for $\mathfrak{o}_N$ invariant models), eigenstates 
expressed in terms of rapidities satisfying the Bethe equations. 
More generally, the expression of Bethe vectors without assuming that the rapidities obey the 
Bethe equations leads to the concept of \textsl{off-shell} Bethe vectors. In that context eigenstates of the transfer matrix will be called \textsl{on-shell} Bethe vectors. Bethe vectors (on-shell and off-shell)
are foundational for understanding the spectrum of integrable models and play 
a central role in calculating physical observables such as correlation functions 
\cite{BoKoIz93}.
The scalar product of (off-shell) Bethe vectors is particularly important, 
as it provides overlaps between Bethe states, allowing the evaluation of matrix 
elements of operators. An example of the use of this scalar product can be found in \cite{Gom24},
where the scalar product of Bethe states in $\mathfrak{gl}_N$ invariant models is used to study the exact overlaps for all integrable two-site boundary states.
The use of scalar products of Bethe vectors to study form factors and correlation functions for $\mathfrak{gl}_2$-related models was implemented in the works of the Lyon group \cite{KMT99, MT00, CM05, KKMST11, KKMST12}. A key tool in the thermodynamic limit  of these correlation functions relies on a determinant form for the form factors, itself inherited from a determinant form for the scalar products of Bethe vectors. Determinant form for form factors of $\mathfrak{gl}_3$ models have been developed in \cite{BPRS12b,PRS2013,PRS2014b}, and in \cite{PRS2016} for form factors of $\mathfrak{gl}_{2|1}$ models.
More generally, such calculations are crucial for studying the dynamics 
and thermodynamics of integrable systems \cite{Slavnov89}.

Depending on the spin chain model one considers, different results have been 
obtained for scalar products of Bethe vectors. 
For $\fg\fl_2$ and $\fg\fl_3$ invariant models, the expression of the 
scalar product as a sum over partition of the Bethe parameter sets has been 
obtained in \cite{IzKoRe87} and \cite{Resh1986}  respectively. The general case 
of $\fg\fl_n$ invariant models has been dealt in \cite{HLPRS2017a}.

The expression for the norm as a determinant of some matrix was suggested by M. Gaudin for the Lieb-Liniger model \cite{Gau72}. Hence the name of Gaudin determinant. Later, the same type of expression was suggested for the XXZ Heisenberg spin chain in \cite{GMW81}. This was subsequently proven in \cite{Kore}
for $\mathfrak{gl}_2$ models, in \cite{Resh1986} for $\mathfrak{gl}_3$ models and in \cite{HLPRS2017b} for 
$\mathfrak{gl}_{n|m}$ models.
An alternative expression for the norms of  eigenstates in $\mathfrak{gl}_n$ based models was calculated using an approach based on the quantized Knizhnik–Zamolodchikov equation \cite{TV94, TV94b, MV}.
An expression of the scalar product as a determinant, when one of the 
Bethe vector is on-shell has been
computed in \cite{Slavnov89} for $\fg\fl_2$ invariant models.

The first accounts on $\mathfrak{o}_n$ invariant integrable models go back to mid-eighties \cite{Resh1985,dVK}, in connection with the $1d$ non-linear Schrodinger equation on symmetric spaces. The general approach of algebraic Bethe ansatz for $\mathfrak{o}_n$ invariant integrable models was done in the nineties \cite{Res91}.
Then, $\fo_{n}$ symmetric models appear in the context of integrable quantum field theories, first with the 
$\fo_{3}$ sigma model \cite{FOZ}, and then for
 $\fo_{2n}$ sigma models \cite{BFK12,BFK13} and $\fo_{2n}$ invariant Gross-Neveu models \cite{BFK16}
 in the context of AdS/CFT correspondence.
In the present paper, we present the case of $\fo_{2n+1}$ invariant models. 
We  compute the scalar product of Bethe vectors in $\fo_{2n+1}$ invariant 
models as a sum over partitions. As for other models, the expression 
makes appear the so-called highest coefficients, and we provide 
recurrence relations and some residue formula for these coefficients. 
We also compute the norm of on-shell Bethe vectors as a Gaudin 
determinant. We show that our results are consistent with the results 
already obtained for $\fg\fl_n$ invariant models, and also with the 
case of $\fo_3$ models, studied in \cite{LPRS2019}.

The plan of the paper is as follows. In section \ref{sect1}, we introduce the algebraic framework we will use in this paper. In section \ref{sect:BV}, we present some general properties of Bethe vectors, such as action formulas, recurrence relations and co-product formula, that have been already established elsewhere 
\cite{LPRS2019,LP2020,LP2021,LPR2024}. 
Section \ref{sect:sum} contains the first main result of the paper: an expression of the scalar product of off-shell Bethe vectors as a sum over partitions of the Bethe parameter sets, see theorem \ref{thm:sum-formula}. The expression makes appear as a building block  the so-called highest coefficient.
In section \ref{sect:recZ}, we use recurrence relations for Bethe vectors to deduce recurrence relations for 
for the full scalar product, see proposition \ref{prop51}, and the 
highest coefficient, see propositions \ref{prop52} and \ref{prop53}. We also present a reduction to $\fg\fl_n$ invariant models, as well as a specialization to the $\mathfrak{o}_3$ case. Section \ref{sect:anal} presents some analytical properties of the highest coefficients, see proposition \ref{prop:residue} and 
is application to the scalar product, see section \ref{sect:resS}. The norm of on-shell Bethe vectors is dealt in section \ref{sect:gaudin}. We show that it takes the form of a Gaudin determinant, see theorem \ref{thm:gaud}. The conclusion is presented in section \ref{conclu}, where possible applications and extensions of our work are elicited. Two appendices are devoted to technical proofs.

\section{Integrable models with $\fo_{2n+1}$ symmetry\label{sect1}}
\paragraph{The Yangian $Y(\fo_{2n+1})$.}
The models we are considering have a $\fo_{2n+1}$ invariance. They are constructed within the Yangian 
$Y(\fo_{2n+1})$, defined through a $(2n+1)\times(2n+1)$ monodromy matrix $T(z)$ obeying the celebrated FRT relations \cite{FRT}:
\begin{equation}\label{rtt}
R(z-w)\,T_1(z)\,T_2(w) 
= T_2(w)\,T_1(z)\,R(z-w)\,,
\end{equation}
where the $R$-matrix for the Yangian $Y(\fo_{2n+1})$ \cite{ZZ79} is defined by
\begin{equation}\label{rmat}
R(z)=\mathbf{I}\otimes\mathbf{I}
+\frac{c}{z}\,\mathbf{P} -
\frac{c}{z+c\kappa}\,\mathbf{Q}\,,\qquad 
\mathbf{P}=\sum_{i,j=-n}^n \ee_{i,j}\otimes \ee_{j,i}
\ \mbox{ and }\ \mathbf{Q}=\sum_{i,j=-n}^n\ee_{i,j}\otimes 
\ee_{-i,-j}
\end{equation}
with $\kappa=n-1/2$ and  the spectral parameter $z$. 
In \eqref{rmat}, we have labeled the indices from $-n$ to $n$, 
and $\ee_{i,j}$ is the elementary matrix with 1 at position $(i,j)$ and
0 elsewhere.  We define the transposition as
\begin{equation}\label{trans}
\big(\ee_{i,j}\big)^{{\rm t}}=\ee_{-j,-i}\,.
\end{equation}
We use the same notation for 
the transposition anti-morphism $(\cdot)^{t}$
 in the algebra of monodromy matrix entries 
determined by this matrix transposition
\begin{equation}
\big(T(z)\big)^{\rm t}
=\sum_{i,j=-n}^n \ee_{i,j}\otimes \big(T_{i,j}(z)\big)^{\rm t}=
\sum_{i,j=-n}^n \big(\ee_{i,j}\big)^{\rm t}\otimes T_{i,j}(z)
=\sum_{i,j=-n}^n \ee_{i,j}\otimes T_{-j,-i}(z)\,.
\end{equation} 

Decomposing the monodromy matrix as 
\begin{equation*}
T(z)=\sum_{i,j=-n}^n\ee_{i,j}\otimes T_{i,j}(z)\,,
\end{equation*}
we get
\begin{equation}\label{comtt}
\begin{split}
  \left[ T_{i,j}(z), T_{k,l}(w) \right] =&  
  \frac{c}{z-w}\Big( T_{k,j}(w)T_{i,l}(z) - T_{k,j}(z) T_{i,l}(w) \Big)\\
  &+\frac{c}{z-w+c\kappa}\sum_{p=-n}^n\Big(\delta_{k,-i} \, T_{p,j}(z)T_{-p,l}(w)-
  \delta_{l,-j}\,  T_{k,-p}(w)T_{i,p}(z)\Big)\,.
  \end{split}
\end{equation}
These relations imply that $T(z)T^t(z+c\kappa)$ is central, and we will impose
\begin{equation}\label{cent}
  T(z)^{\rm t}\cdot T(z+c\kappa)=  T(z+c\kappa)\cdot T(z)^{\rm t} = 
  \mathbf{I}\,.
\end{equation}
Indeed, starting from any model, a simple rescaling  of the monodromy matrix by a function $f(z)$, $T(z) \to f(z)T(z)$, will ensure that the condition \eqref{cent} is fulfilled.

\paragraph{The generalized models.}
From the monodromy matrix, one introduces an algebraic transfer matrix
\begin{equation}\label{eq:transfer}
\mathcal{T}(z)=\sum_{j=-n}^n T_{j,j}(z).
\end{equation}
The FRT relation ensures that $[\mathcal{T}(z)\,,\, \mathcal{T}(w)]=0$, 
so that, upon expansion in the spectral parameter, 
the transfer matrix generates an 
Abelian subalgebra of the Yangian. 

To get a physical model, one needs to specify a representation of the 
Yangian algebra. 
Typically, one takes a tensor product of evaluation representations to get a 
spin chain on a periodic one-dimensional lattice. 
However, most of the calculations concerning the Bethe vectors can be 
performed at the algebraic level, leading to the notion of generalized models. In 
these models, one solely imposes the conditions
\begin{equation}\label{eq:hw}
T_{i,i}(z)|0\rangle = \lambda_i(z)|0\rangle \quad\text{and}\quad 
T_{i,j}(z)|0\rangle =0, \ -n\leq j<i\leq n\,,
\end{equation}
where $|0\rangle$ 
is the so-called vacuum state and $\lambda_i(z)$ are some functions of one variable. The functions $\lambda_i(z)$, $i=1,2,...,n$ are arbitrary free functions, while the remaining ones, due to the relation \eqref{cent}, are constrained by the relation
\begin{equation}\label{lam}
\lambda_{-j}(z)=\frac{1}{\lambda_{j}(z_j)}\ \prod_{s=j+1}
^n\frac{\lambda_s(z_{s-1})}{\lambda_s(z_s)},\qquad j=0,1,\ldots,n\,,
\end{equation}
where $z_s=z-c(s-1/2)$, $s=0,1,\ldots n$. 

The 'real' physical models are obtained by giving an explicit form 
for the functions $\lambda_i(z)$. For example, one can consider 
\begin{equation*}
\begin{split}
\lambda_{-n}(z)= a(z)\left(1+\frac{c}z\right)^L\,;\quad 
\lambda_j(z)&=a(z)\,,\ -n<j<n\,;\quad 
\lambda_{n}(z)=a(z)\left(1-\frac{c}{z+c\kappa}\right)^L\,,\\
a(z)\,a(z-c\kappa)&=\left(\frac{z^2}{\big(z+c\big)
\big(z-c\big)}\right)^L\,,
\end{split}
\end{equation*}
where $a(z)$ has been fixed in such a way that
\eqref{cent} and \eqref{lam} are satisfied. 
This choice of $\lambda_j(z)$ corresponds to a spin 
chain model of $L$ sites, each of them carrying a fundamental ($(2n+1)$-dimensional )
representation of the $\fo_{2n+1}$ Lie algebra. In this spin chain, 
the monodromy matrix is realized as a product of $R$ matrices:
$$
T_0(z)={a(z)^L}\,R_{0L}(z)\cdots R_{01}(z).
$$ 

Let us insist that this is just an example of the models we deal with. 
In the following, we will keep the functions $\lambda_i(z)$ free, up to the relation \eqref{lam}, and the 
monodromy matrix as algebraic, just obeying the relations 
\eqref{comtt}, \eqref{cent} and \eqref{eq:hw}. 

\paragraph{Notations.}
We will use different sets of parameters, indexed by an integer 
$s$ running from 0 to $n-1$ and called the color of the set: 
$\bar u^{(s)}=\{u^{(s)}_1,\dots,u^{(s)}_{{r_s}}\}$. As a convention,
 sets (and subsets) will be always noted with a bar, as in the 
 preceding example. The cardinality of these sets will be noted
$|\bar u^{(s)}|=r_s$. Collections of such sets will be noted as 
$\{\bar u^{(i)},\bar u^{(i+1)},...,\bar u^{(j)}\}=
\{\bar u^{(s)}\}_{s=i}^{j}$  
and the full collection of sets
as $\bar u=\{\bar u^{(0)},\bar u^{(1)},...,\bar u^{(n-1)}\}=
\{\bar u^{(s)}\}_{s=0}^{n-1}$. 
As an additional notation, $\bv_k$ will be the subset complementary to the element $v_k$ in the set $\bv$: $\bv^{(s)}_k = \bv^{(s)}\setminus \{v^{(s)}_k\}$.

We will also use sums over partitions of these sets in two or 
three subsets, e.g. $\bar u^{s}\vdash\{\bar u^{s}_{\so},\bar u^{s}_{\st}\}$, 
where $\bar u^{s}_{\so}$ and $\bar u^{s}_{\st}$ are disjoint 
(possibly empty) subsets such that 
$\bar u^{s}_{\so}\cup\bar u^{s}_{\st}=\bar u^{s}$. Let us stress that the subsets 
may be empty, although 
in general this does not occur in a partition.
For example,
\begin{equation*}
    \sum_{\bu = \{u_1, u_2\} \vdash\{\bar u_{\so},\bar u_{\st}\}} G(\bu_{\so}, \bu_{\st}) = 
    G(\varnothing, \bu) + G(u_1, u_2) + G(u_2, u_1) + G(\bu, 
    \varnothing).
\end{equation*}

We define the functions
\begin{equation}\label{eq:functions}
\begin{split}
&f(u,v)=\frac{u-v+c}{u-v}\ ,\quad \ff(u,v)=\frac{u-v+c/2}{u-v}\ ,
\quad g(u,v)=\frac{c}{u-v}\ ,\\
&h(u,v)=\frac{f(u,v)}{g(u,v)}=\frac{u-v+c}{c}\ ,\quad \gamma_s(u,v)=\begin{cases}\displaystyle\ \ff(u,v)\qquad 
\text{when } s=0\,,\\[1ex] 
\displaystyle \frac{g(u,v)}{h(v,u)}\qquad \text{otherwise}\,,\end{cases}\quad\\ 
&\alpha_s(u)=\frac{\lambda_{s}(u)}{\lambda_{s+1}(u)}\,.
\end{split}
\end{equation}

To lighten the presentation of the results, we will use the following 
convention. For any function depending on
one or two variables, if a set appears as a variable, then one has 
to consider the product of this function at all element of the set. 
For instance for a set $\bar u^{(s)}$ of cardinality $r_s$, 
$f(\bar u^{(s)},\bv)=\prod_{\ell=1}^{r_s}
f(u^{(s)}_{\ell},\bv)$. 
As a rule, we will also set $f(\varnothing,\bv)=1$.

\paragraph{Bethe vectors.} Usually, Bethe vectors are 
defined as eigenvectors of the transfer matrix
\begin{equation*}
\begin{split}
\mathcal{T}(z)\,\BB(\bar u) &= \tau(z;\bar u)\,\BB(\bar u)\,,\\
 \tau(z;\bar u)&= \lambda_0(z)f(\bar u^{(0)},z_0)f(z,\bar u^{(0)})\ +\\
 &+\sum_{s=1}^n \lambda_{s}(z)f(\bar u^{(s)},z)f(z,\bar u^{(s-1)})  +
 \lambda_{-s}(z)f(\bar u^{(s-1)},z_{s-1})f(z_s,\bar u^{(s)}) 
\end{split}
\end{equation*}
provided the Bethe equations are obeyed
\begin{equation}\label{eq:BAE}
\begin{split}
\alpha_0(u^{(0)}_k) =& \frac{\ff(u^{(0)}_k,\bar u^{(0)}_k)\,f(\bar u^{(1)},u^{(0)}_k)}
{\ff(\bar u^{(0)}_k, u^{(0)}_k)}\,,\quad\quad k=1,...,r_0\,,\\
\alpha_s(u^{(s)}_k) =& \frac{f(u^{(s)}_k,\bar u^{(s)}_k)\,f(\bar u^{(s+1)},u^{(s)}_k)}
{f(\bar u^{(s)}_k, u^{(s)}_k)\,f( u^{(s)}_k,\bar u^{(s-1)})}\,,\quad k=1,...,r_s\,,\ 
s=1,...,n-1.
\end{split}
\end{equation}
We will call these vectors \textit{on-shell Bethe vectors} because their Bethe 
parameters obey the Bethe equations \eqref{eq:BAE}.

If we do not require the Bethe equations to be satisfied, we will call the 
corresponding vector an \textit{off-shell Bethe vector}. 
Bethe vectors (on-shell or off-shell) are polynomials of the 
monodromy matrix entries acting on the vacuum vector state:
\begin{equation}\label{eq:preBV}
\BB(\bar u) = \cP\big(\{T_{i,j}(u^{(s)}_k),\ -n\leq i < j\leq n,\ 1\leq k \leq r_s, \ 
0\leq s\leq n-1\}\big)\,|0\rangle \equiv \cB(\bar u)\,|0\rangle\,,
\end{equation}
The polynomials which defines the 
off-shell Bethe vectors in \r{eq:preBV}
are not arbitrary. They are constrained by requiring that the Bethe vector $\BB(\bar u)$ becomes an eigenvector of the transfer matrix when the Bethe equations are satisfied. In fact, these constraints are not sufficient to uniquely define the Bethe vectors, a coproduct property is needed that we introduce later in section \ref{sect:coprod}. A precise definition of Bethe vectors can be found in \cite{LP2021}.
In \eqref{eq:preBV}, $\cB(\bar u)$ is called a \textit{pre-Bethe vector}.

Besides the transposition \r{trans} we define 
a usual matrix transposition 
\begin{equation}\label{eq: anti morph}
\big(\ee_{i,j}\big)^{\prime}=\ee_{j,i}\,.
\end{equation}
As for the transposition \r{trans}, we use the same notation for 
the transposition anti-morphism $(\cdot)^{\prime}$
 in the algebra of monodromy matrix entries 
determined by this matrix transposition
\begin{equation}\label{a-mor}
\big(T(z)\big)^{\prime}
=\sum_{i,j=-n}^n \ee_{i,j}\otimes \big(T_{i,j}(z)\big)^{\prime}=
\sum_{i,j=-n}^n \big(\ee_{i,j}\big)^{\prime}\otimes T_{i,j}(z)
=\sum_{i,j=-n}^n \ee_{i,j}\otimes T_{j,i}(z)\,.
\end{equation} 
Being extended to the vacuum vectors  
\begin{equation}\label{eq:vac}
 \langle0| =|0\rangle^{\prime}\,,\qquad |0\rangle=\langle0|^{\prime}\,,
\end{equation}
this transposition 
anti-morphism allows to define 
 the dual Bethe vectors as follows
\begin{equation}\label{dual:BV}
\CC(\bar u^{(0)}, \dots, \bar u^{(n-1)})=\Big(\BB(\bar u^{(0)}, 
\dots, \bar u^{(n-1)})\Big)^{\prime}\,, 
\end{equation}
where  according to \r{a-mor}
$\big(T_{i,j}(u)\big)^{\prime} = T_{j,i}(u)$. 
Indeed, the anti-morphism transposition $(\cdot)^{\prime}$ maps  Yangian left highest weight representations  to right highest weight representations. The relation \eqref{eq:vac} just states that the left highest weight vector $|0\rangle$  can be mapped to a right highest weight vector $\langle0|$. Since there is an isomorphism between left and right representations, the condition \eqref{eq:vac} does not impose any constraint on the type of representations.
As a consequence, all
 the definitions given above also apply to dual 
Bethe vectors. 

\section{Properties of Bethe vectors\label{sect:BV}}

In this section, we present different properties of off-shell Bethe vectors (BVs), 
that will be needed for the calculation of their scalar products.

\subsection{Action formula on BVs}

To describe the action of the monodromy entry $T_{i,j}(z)$ on the 
off-shell Bethe vector $\BB(\bu)$  we define the extended sets 
 $\bar w^{(s)}=\bar u^{(s)}\cup\{z,z_s\}$ for $s=0,1,\ldots,n-1$
 with $z_s=z-c(s-1/2)$.
The action formula has been given  in equation (3.5) of
\cite{LP2020} as a sum over partitions $\bar w^{(s)}\vdash \{\bar 
w^{(s)}_{\so},\bar w^{(s)}_{\st},\bar w^{(s)}_{\sth}\}$, with  cardinalities
\begin{subequations}
\label{eq:part-tot-TijB}
\begin{equation}
\big|\bar w^{(s)}_{\so}\big|=\begin{cases} 2, &\ s<i\leq n, \\  
1, & -s\leq i\leq s, \\  0, & -n\leq i<-s, \end{cases}
\quad \text{ and }\quad \big|\bar w^{(s)}_{\sth}\big|=
\begin{cases} 0, &\ s< j\leq n, \\  1,& -s\leq j\leq s, \\  2, & -n\leq j< -s, 
\end{cases}
\label{eq:part-TijB} 
\end{equation}
\begin{equation}
 \big|\bar w^{(s)}_{\so}\big|+\big|\bar w^{(s)}_{\st}\big|+
 \big|\bar w^{(s)}_{\sth}\big|=\big|\bar w^{(s)}\big|.
\label{eq:compa-partitions}
\end{equation}
\end{subequations}
\begin{rmk}
The condition \eqref{eq:compa-partitions} should be automatically satisfied 
since we have a partition. However, in some particular cases (depending on 
$i$, $j$ and the cardinalities of $\bar u$), the conditions  \eqref{eq:part-TijB} 
may contradict this condition: in that case, the corresponding term should 
be discarded, leading to a vanishing of the action formula for the 
corresponding $T_{i,j}(z)$. An example of such case is given below, 
for the reduction to the $\fg\fl_{n}$ case.
\end{rmk}

For two collections of sets $\bar{x} = \left\{\bar{x}^{(0)}, \bar{x}^{(1)}, \ldots, \bar{x}^{(n-1)} \right\}$ and $\bar{y} = \left\{\bar{y}^{(0)}, \bar{y}^{(1)}, \ldots, \bar{y}^{(n-1)} \right\}$ 
we define the 
function 
\begin{equation}\label{eq:defOmega}
\Omega(\bar{x}|\bar{y})=\prod_{s=0}^{n-1}  
\gamma_s(\bar{x}^{(s)},\bar{y}^{(s)})\ 
\frac{ h(\bar{y}^{(s+1)},\bar{x}^{(s)})}{ g(\bar{x}^{(s+1)},\bar{y}^{(s)})}\,.
\end{equation}

We also define the functions $\Phi_{i,j}(\bw)$
\begin{equation}\label{eq:Phi-ij}
\Phi_{i,j}(\bw)=-\ \sigma_i\sigma_{-j}\,
\frac{g(z_1,\bar u^{(0)})}{\kappa\,h(z,\bar u^{(0)})}\ 
\Omega(\bw_{\so}|\bw_{\st})\,\Omega(\bw_{\st}|\bw_{\sth})\,
\Omega(\bw_{\so}|\bw_{\sth})\,,
\end{equation}
where 
\begin{equation}\label{eq:sigma}
\sigma_i=\begin{cases} -1, &\text{if } i<1,\\ 1,  &\text{if } i\geq 1.\end{cases}
\end{equation}
Let us stress that $\Phi_{ij}(\bar w)$ does not depend on the functions 
$\lambda_s$. 
Note also that although the expression of $\Phi_{ij}(\bar w)$ does not 
seem to depend on $i$ and $j$, this dependence is hidden in the 
cardinalities of the partitions, as detailed in \eqref{eq:part-tot-TijB}.
The function $\Phi_{i,j}(\bw)$ depends on  an additional 
boundary 
set $\bar w^{(n)}=\{z,z_n\}$ with a fixed partition 
\begin{equation}
\label{eq:part-sup-TijB}
\bar w^{(n)}_{\so}=\{z_n\}\,, \qquad \bar w^{(n)}_{\st}=
\varnothing\,, \qquad\bar w^{(n)}_{\sth}=\{z\} \,.
\end{equation}

The action of the monodromy matrix entries on BVs reads \cite{LP2020}
\begin{equation}\label{eq:TijB}
{T_{i,j}(z)\cdot \BB(\bar u)}= {\lambda_n(z)} \sum_{\rm part} \left(\prod_{s=0}
^{n-1}\alpha_s(\bar w^{(s)}_{\sth})\right)\Phi_{i,j}(\bar w) \,\BB(\bar w_{\st})\, ,
\end{equation}
where the sum goes over partition 
$\sw{s}\vdash
\{\sw{s}_{\so},\sw{s}_{\st},\sw{s}_{\sth}\}$ with the subsets
cardinalities given by 
\r{eq:part-tot-TijB}.

Relying on the Yangian $Y(\fo_{2n+1})$, we assume that the monodromy matrix has the following dependence  on the formal 
parameter $z$ 
\begin{equation}\label{zm1}
T_{i,j}(z)=\delta_{i,j}+\sum_{m\geq 0}T_{i,j}[m]\,(z/c)^{-m-1}
\end{equation}
and we define the
zero modes $\sT_{i,j}\equiv T_{i,j}[0]$ as
\begin{equation}\label{zm2}
\sT_{i,j}=\mathop{\rm lim}_{z\to\infty}\ \frac{z}{c}\ \Big(T_{i,j}(z)-\delta_{i,j}\Big)\,.
\end{equation} 
These zero modes satisfy the commutation relations of the 
$\fo_{2n+1}$ algebra 
\begin{equation}\label{zm3}
[\sT_{i,j},\sT_{k,l}]=\Big(\delta_{i,l}\,\sT_{k,j}-\delta_{j,k}\,\sT_{i,l}\Big)-
\Big(\delta_{j,-l}\,\sT_{k,-i}-\delta_{i,-k}\,\sT_{-j,l}\Big).
\end{equation}
Let $\sh_s$ for $s=0,1,\ldots,n-1$ be the operators 
\begin{equation}\label{zm4}
\sh_s=\sum_{i=s+1}^n\Big(T_{i,i}[0]-\lambda_i[0]\Big)
\end{equation}
such that $\sh_s\,\rvec=0$ and $\lvec\,\sh_s=0$. The adjoint action 
of the operators $\sh_s$ on the monodromy entry $T_{i,j}(z)$ reads
\begin{equation}\label{adj-t}
[\sh_s,T_{i,j}(z)]=\sum_{\ell=s+1}^n\Big(\delta_{\ell,j}-\delta_{\ell,i}+\delta_{\ell,-i}-
\delta_{\ell,-j}\Big)\ T_{i,j}(z)\,.
\end{equation}
For $i<j$, relation \eqref{adj-t} always yields a non-negative eigenvalue 
equal to either $0$, $1$ or $2$ depending 
on relations between the indices $s$, $i$, and $j$.

\begin{prop}\label{dzmac}
The operators $\sh_s$ being applied to the BVs and dual BVs 
measure the cardinalities of their Bethe parameters:
\begin{equation}\label{zm5}
\sh_s\cdot\BB(\bu)=r_s\,\BB(\bu),\qquad \CC(\bu)\cdot\sh_s=r_s\,\CC(\bu)\,,\quad s=0,...,n-1.
\end{equation}
\end{prop}
\proof 
This proposition can be proved by induction on the total cardinality $|\bu|$, using \r{adj-t}
 and the recurrence relations for the Bethe vectors \r{eq:recBBgen}
described in the next section. We will prove only the first equation in \r{zm5} 
since the second one results from the application of the anti-morphism \r{dual:BV}
to the first. The base of the induction is the case when all sets 
are empty: according to the definition \r{zm4}
$\sh_s\cdot\BB(\varnothing)=\sh_s\cdot\rvec=0$ for $s=0,1,\ldots,n-1$.
Assume now that the first equation in \r{zm5} is valid for $|\bu|=r$ and any $s$, and apply an operator 
$\sh_s$ to both sides of the recurrence relation \r{eq:recBBgen}. 
Using \r{adj-t} and the description of the cardinalities \r{eq:part-recBgen}
we find that 
\begin{equation*}
\sh_s\cdot\BB(\bu,z^{(\ell)})=(r_s+\delta_{s,\ell})\,\BB(\bu,z^{(\ell)})\,.
\end{equation*}
Since $|\{\bu,z^{(\ell)}\}|=r+1$, this proves the induction and  finishes the proof of the proposition~\ref{dzmac}.\qed

\subsection{Recurrence relations for BVs\label{sect:rec}}

To lighten the presentation we introduce notation 
\begin{equation}\label{eq:def-z^ell}
\big(\bu,z^{(\ell)}\big)=\big(\big\{\su{s}\big\}_{s=0}^{\ell-1},\{\su{\ell},z\},
\big\{\su{s}\big\}_{s=\ell+1}^{n-1}\big)\,,
\end{equation}
for any fixed $\ell=0,1,\ldots,n-1$. 
It can be shown that the BVs in the $\mathfrak{o}_{2n+1}$ 
invariant models obey several types of recurrence relations \cite{LPR2024}. 
For a given $\ell$ the recurrence relations relevant for this paper 
take the general form\footnote{In \cite{LPR2024} another type 
recurrence relations was also considered when the set $\su{\ell}$ 
is extended by the shifted parameter $z_\ell$. In this paper we will 
not use these so-called shifted recurrence relations.}
\begin{equation}\label{eq:recBBgen}
\BB(\bu,z^{(\ell)})
= \sum_{i=-n}^{\ell}\sum_{j=\ell+1}^{n} \sum_{\rm part}\sk{\prod_{s=\ell+1}
^{j-1}\alpha_s(\bar u_{\sth}^{(s)})}\
\Psi^{(\ell)}_{i,j}(\bar u, z)\,\frac{T_{i,j}(z)\cdot \BB(\bar u_{\st})}
{\lambda_{\ell+1}(z)}\,,
\end{equation}
where the sum is on partition of $\bar u^{(s)}\vdash \{\bar u^{(s)}_{\so},\bar 
u^{(s)}_{\st},\bar u^{(s)}_{\sth}\}$, with  cardinalities depending on $i$ and $j$ 
as follows
\begin{equation}\label{eq:part-recBgen}
\begin{aligned}
&\mb{for} 0\leq s<\ell : \qquad 
&&|\bar u^{(s)}_{\so}|=
\begin{cases}2\,, &i< -s\leq0\,, \\  1\,, & -s\leq i\leq s\,, \\ 0\,,\ & s< i \,,\end{cases} 
\qquad 
&&|\bar u^{(s)}_{\sth}|=0\,,
\\[1ex]
&\mb{for} s= \ell : \quad 
&&|\bar u^{(\ell)}_{\so}|=
\begin{cases} 1,  & i<-\ell, \\ 0, & -\ell\leq i, \end{cases} 
\qquad&& |\bar u^{(\ell)}_{\sth}|=0\,,
\\[1ex]
&\mb{for} \ell <s\leq n-1: \quad 
&&|\bar u^{(s)}_{\so}|=
\begin{cases} 1,  & i< -s, \\ 0, & -s\leq i, \end{cases}
\qquad &&|\bar u^{(s)}_{\sth}|=
\begin{cases} 1, & s< j, \\ 0,  & j\leq s. \end{cases} 
\end{aligned}
\end{equation}
The function
$\Psi^{(\ell)}_{i,j}(\bar u, z)$ is a rational function that does not depend on the 
functions $\alpha_s$.
Its explicit expression depends on the indices $i$, $j$ and $\ell$ as follows.
\paragraph{In the case $\boldsymbol{\ell>0}$,} they take the form
\begin{equation}\label{eq:Psi-gen}
\Psi^{(\ell)}_{i,j}(\bar u, z)= \sigma_{i+1}\,\frac{
g(z,\bar u^{(\ell-1)}_{\so})h(\bar u^{(\ell)}_{\so},z)g(\bar u^{(\ell+1)}_{\sth},z)}
{g(z,\bar u^{(\ell-1)})\,h(z,\bar u^{(\ell)})\,h(\bar u^{(\ell)},z)\,g(\bar u^{(\ell+1)},z)}
\ \Omega(\bu_{\so}|\bu_{\st})\,\Omega(\bu_{{\so},{\st}}|\bu_{\sth}),
\end{equation}
where notation $\bu_{{\so},{\st}}$ means the union 
$\bu_{{\so}}\cup\bu_{{\st}}$. According to our convention 
for the products of the rational functions 
$\Omega(\bu_{{\so}, {\st}},\bu_{\sth})=
\Omega(\bu_{{\so}}|\bu_{\sth})\Omega(\bu_{{\st}}|\bu_{\sth})$.
Note that for $\ell>0$, we must have $n\geq2$.
\paragraph{In the case $\boldsymbol{\ell=0}$,} we have
\begin{equation}\label{eq:Psi-0}
\Psi^{(0)}_{i,j}(\bar u, z)=\sigma_{i+1}
\frac{g(z_0,\bar u^{(0)}_{\so})\, g(\bar u^{(1)}_{\sth},z) }
{g(z_0,\bar u^{(0)})\,h(z,\su{0})\,g(\bar u^{(1)},z)}
\ \Omega(\bu_{\so}|\bu_{\st})\,\Omega(\bu_{{\so},{\st}}|\bu_{\sth})
\end{equation}
with  the cardinalities
\begin{equation}\label{r1B24}
\begin{split}
&|\bar u^{(0)}_{\so}|=\begin{cases} 1, \quad\text{if }i<0,\\ 0, 
\quad\text{otherwise}, \end{cases}\qquad 
|\bar u^{(0)}_{\sth}|=0\,,\\
&|\bar u^{(s)}_{\so}|=\begin{cases} 1, \quad\text{if }s<-i, \\ 0, 
\quad\text{otherwise}, \end{cases}\qquad 
|\bar u^{(s)}_{\sth}|=\begin{cases} 1, \quad\text{if }s<j, \\ 0, 
\quad\text{otherwise}, \end{cases}\quad s=1,\ldots,n-1\,.
\end{split}
\end{equation}
The recurrence relation \r{eq:recBBgen} with the function \r{eq:Psi-0}
is valid also 
in the case $n=1$ and coincides with 
 the recurrence relation given in \cite{LPRS2019}.

\subsection{Recursion formula for dual BVs}

From the recurrence relations \eqref{eq:recBBgen}, applying 
the transposition anti-morphism \r{eq: anti morph}, we get recurrence relations for dual BVs.
For a given $\ell=0,1,...,n-1$, they take the general form
\begin{equation}\label{eq:recCCgen}
\CC(\bv,z^{(\ell)}) = \sum_{j=-n}^{\ell}\sum_{i=\ell+1}^{n} \sum_{\rm part} 
\Big(\prod_{s=\ell+1}^{i-1}\alpha_s(\bar v_{\sth}^{(s)})\Big)\
\Psi^{(\ell)}_{j,i}(\bar v, z)\,\,\frac{\CC(\bar v_{\st})\cdot T_{i,j}(z)}
{\lambda_{\ell+1}(z)}\,,
\end{equation}
with cardinalities
\begin{equation}\label{eq:part-recCgen}
\begin{aligned}
&\mb{for} 0\leq s<\ell : \qquad 
&&|\bar v^{(s)}_{\so}|=
\begin{cases}2\,, &j< -s\leq0\,, \\  1\,, & -s\leq j\leq s\,, \\ 0\,,\ & s< j \,,\end{cases} 
\qquad 
&&|\bar v^{(s)}_{\sth}|=0\,,
\\[1ex]
&\mb{for} s= \ell : \quad 
&&|\bar v^{(\ell)}_{\so}|=
\begin{cases} 1,  & j<-\ell, \\ 0, & -\ell\leq j, \end{cases} 
\qquad&& |\bar v^{(\ell)}_{\sth}|=0\,,
\\[1ex]
&\mb{for} \ell <s\leq n-1: \quad 
&&|\bar v^{(s)}_{\so}|=
\begin{cases} 1,  & j< -s, \\ 0, & -s\leq j, \end{cases}
\qquad &&|\bar v^{(s)}_{\sth}|=
\begin{cases} 1, & s< i, \\ 0,  & i\leq s. \end{cases} 
\end{aligned}
\end{equation}
The functions $\Psi^{(\ell)}_{j,i}(\bar v, z)$ are deduced from 
the expressions \eqref{eq:Psi-gen} and \eqref{eq:Psi-0}.

\subsection{Coproduct property for BVs\label{sect:coprod}}
We remind the well-known expression for the standard 
coproduct for the monodromy matrix:
\begin{equation*}
\Delta\Big( T_{i,j}(u)\Big) = 
\sum_{k} T_{k,j}(u)\otimes T_{i,k}(u) = \sum_{k} T^{[1]}_{k,j}(u)\, T^{[2]}_{i,k}(u) \,.
\end{equation*}
Since pre-Bethe vectors $\cB(\bu)$ are certain polynomials of the monodromy matrix entries $T_{i,j}(u^{(s)}_k)$  one can address 
the question to calculate their coproduct. In general, this
is a quite non-trivial combinatorial problem which was solved 
for $\fg\fl_2$ invariant BVs in the 80's \cite{Kore} and lately 
for $\fg\fl_n$ invariant BVs in \cite{TV94}. In the latter paper 
a trace formula for the pre-Bethe vectors was used to 
prove the coproduct properties of  $\fg\fl_n$ invariant BVs. 
Lately on, an alternative approach to obtain the presentation 
for the $U_q(\fg\fl_n)$ invariant pre-Bethe vectors in terms of 
the Cartan-Weyl generators of the quantum affine algebra was 
proposed \cite{KhP08}. 
Both approaches, the trace formula and the Cartan-Weyl presentation, 
produce slightly different expressions for the pre-Bethe vectors.
Nevertheless these expressions are different only by terms 
which are  annihilated on the vacuum state, but 
 the calculation of the coproduct of BVs in the Cartan-Weyl approach 
 is rather simple. It uses the relation between the standard and Drinfeld 
coproducts of the simple root Cartan-Weyl generators in quantum 
affine algebra proved in \cite{EKhP07}.

An extension of the Cartan-Weyl approach for the description of the 
off-shell Bethe vectors in $\fg\fl(m|n)$ invariant models was achieved in 
\cite{HLPRS17} and for $\fo_{2n+1}$ invariant models in \cite{LP2020}.
In order to develop this description one has to replace the Yangian 
by its double. 
 In the current presentation for the Yangian double $DY(\mathfrak{o}_{2n+1})$, 
 the Bethe 
vectors take the form of a projection of currents:
\begin{equation}\label{offBV}
\BB(\bu)=\cB(\bar u)\rvec=\cP^+\Big(\FF(\bu)\Big)|0\rangle\,
\end{equation}
where $\cP^+$ is a projection on the intersection of  different  
Borel subalgebras in the Yangian double  and 
$\FF(\bu)$ is a product of  Cartan-Weyl generators in a certain order. 
For more details on the projection method in $\mathfrak{o}_{2n+1}$ models, 
see e.g. \cite{LP2020}.
In this framework, the  coproduct of pre-Bethe vectors can be computed assuming that 
there is the same relation between standard and Drinfeld coproduct 
for the Yangian doubles as for the quantum affine algebras
and using the Drinfeld coproduct of the currents. One gets
\begin{equation}\label{co-pr-bv}
\Delta\Big(\cB(\bar u)\Big)\sim  \sum_{\rm part}
 \Omega(\bar u_{\st}|\bar u_{\so})\, \cB(\bar u_{\so})\otimes 
 \cB(\bar u_{\st})\,\prod_{s=0}^{n-1} \Big(\II\otimes 
 T_{ss}(z)(\bar u^{(s)}_{\so})\, T_{s+1,s+1}(z)(\bar u^{(s)}_{\so})^{-1}\Big) \, ,
\end{equation}
where the equivalence $\sim$ means an equality modulo terms 
which are annihilated on the tensor product of the vacuum states 
$\rvec\otimes\rvec$. The sum runs over partitions 
$\bar u^{(s)}\vdash \{\bar u^{(s)}_{\so},\bar u^{(s)}_{\st}\}$. 
Since, as far as we know,  the trace formula for $\fo_{2n+1}$ invariant
Bethe vector is not known, the Cartan-Weyl generators approach is 
a unique way to find formulas for the pre-Bethe vectors and their coproducts 
in $\fo_{2n+1}$ invariant integrable models.

In the following we will use the formula \r{co-pr-bv}
in a composite model, where the 
monodromy matrix is splitted into 
two submodel monodromy matrices: $\Delta(T(u))=T^{[2]}(u)\, T^{[1]}(u)$. 
Correspondingly the eigenvalues $\lambda_s(u)$ factorize in the same way, and for instance $\alpha_s(u)=\alpha_s^{[1]}(u)\alpha_s^{[2]}(u)$. 
The coproduct property allows to compute the Bethe vector 
of the composite model as the product of Bethe vectors corresponding to the
 submodels:
\begin{equation}\label{eq:copBB}
\Delta\Big(\BB(\bar u)\Big)= \sum_{\rm part} \Omega(\bar u_{\st}|\bar u_{\so})\, 
\BB^{[1]}(\bar u_{\so})\, \BB^{[2]}(\bar u_{\st})\,\prod_{s=0}^{n-1} 
\alpha_s^{[2]}(\bar u^{(s)}_{\so}) \,,
\end{equation}
where $\BB^{[1]}(\bar u_{\so})$ and $\BB^{[2]}(\bar u_{\st})$ are BVs for the 
submodels based on $T^{[1]}(u)$ and $T^{[2]}(u)$ respectively.
Note that the Bethe equations \r{eq:BAE} can be rewritten in the form
\begin{equation}
\prod_{s=0}^{n-1} \alpha_s(\bar u^{(s)}_{\so}) = \frac{\Omega(\bar u_{\so}|\bar u_{\st})}{\Omega(\bar u_{\st}|\bar u_{\so})}, \quad \text{for any partitions}\quad \bar u^{(s)}\vdash \{\bar u^{(s)}_{\so},\bar u^{(s)}_{\st}\},\quad s = 0, \ldots, n-1.
\end{equation}
This implies that for on-shell BVs, there is an alternative form to the coproduct formula:
\begin{equation}
\Delta\Big(\BB(\bar u)\Big)= \sum_{\rm part} \frac{\Omega(\bar u_{\so}|\bar u_{\st})\, 
\BB^{[1]}(\bar u_{\so})\, \BB^{[2]}(\bar u_{\st})}{\prod_{s=0}^{n-1} 
\alpha_s^{[1]}(\bar u^{(s)}_{\so})}  
=\frac{\Delta^{op}\Big(\BB(\bar u)\Big)}{\prod_{s=0}^{n-1} 
\alpha_s^{[1]}(\bar u^{(s)})}\,,
\end{equation}
where $\Delta^{\rm op}=\sigma\circ\Delta$ with 
$\sigma(A^{[1]}\,B^{[2]})=A^{[2]}\,B^{[1]}$.

\section{Sum formula for the scalar products in $Y(\fo_{2n+1})$ models\label{sect:sum}}
The sum formula for the scalar product of BVs in $Y(\fo_{2n+1})$ models expresses the scalar product as a sum over partition of  rational functions and product of $\alpha_s$ functions. It has been proven in the case of $Y(\fg\fl_2)$ models in \cite{IK1987}, and then generalized in \cite{Resh1986} for $Y(\fg\fl_3)$ and in \cite{HLPRS2017a,HLPRS2017c} for $Y(\fg\fl_{m|n})$ and $U_q( \widehat{\fg\fl}_n)$ models. The goal of this section is to prove it for $Y(\fo_{2n+1})$ models and is summarized in the theorem \ref{thm:sum-formula}.

As usual in integrable systems, we  define the scalar product of the Bethe vectors $\BB(\bar v)$ and $\BB(\bar u)$ as $S(\bar v|\bar u) = \CC(\bar v)\,\BB(\bar u)$.
\begin{lemma}\label{lem:sym}
\begin{enumerate}[(i)]
\item The scalar product $S(\bar v|\bar u)$ is symmetric in the exchange $\bar u\ \leftrightarrow\ \bar v$.
\item It is also invariant under any permutation within the set $\bar u^{(s)}$ and within the set $\bar v^{(s)}$.
\item $S(\bar v|\bar u)=0$ whenever there is at least one color $s$ such that $|\bar v^{(s)}|\neq |\bar u^{(s)}|$.
\end{enumerate}
\end{lemma}
\proof Since $S(\bar v|\bar u)$ is a scalar, we have 
\begin{equation*}
S(\bar v|\bar u)=S(\bar v|\bar u)^{{\prime}}=
\BB(\bar u)^{{\prime}}\,\CC(\bar v)^{{\prime}}= 
\CC(\bar u)\,\BB(\bar v)=S(\bar u|\bar v).
\end{equation*}
Hence the first claim. The second one is a direct consequence of the same property for BVs themselves.
Finally, using operators $\sh_s$ defined by \r{zm4} from 
\begin{equation*}
\CC(\bar v)\,\big(\sh_s\,\BB(\bar u)\big)=
|\su{s}|\,\CC(\bar v)\,\BB(\bar u)=\big(\CC(\bar v)\,\sh_s\big)\,\BB(\bar u)=|\sv{s}|\,\CC(\bar v)\,\BB(\bar u)\,,
\end{equation*}
 we get the last claim.
\qed

\begin{lemma}\label{lem42}
The scalar product $S(\bar v|\bar u) $ depends on the functions $\alpha_s(u_j^{(s)})$ and $\alpha_s(v_k^{(s)})$, $1\leq j,k\leq r_s$, $s=0,\dots,n-1$, but not
on the functions $\alpha_s( u_j^{(\t)})$ or $\alpha_s( v_k^{(\t)})$, $\t\neq s$.
Moreover, each Bethe parameter $u_j^{(s)}$ or $v_k^{(s)}$ occurs at most once in the function $\alpha_s$.
\end{lemma}
The proof of this lemma can be found in appendix~\ref{ApA}.\qed

\begin{prop}
The scalar product $S(\bar v|\bar u) $ can be written as
\begin{equation}\label{eq:scal1}
S(\bar v|\bar u)=\sum_{part} W(\bar v_{\so},\bar v_{\st}|\bar u_{\so},\bar u_{\st})\, \prod_{s=0}^{n-1}\alpha_s(\bar v^{(s)}_{\so})\,\alpha_s(\bar u^{(s)}_{\st})\,,
\end{equation}
where the sum is taken over partitions $\bar u^{(s)}\vdash \{\bar u^{(s)}_{\so},\bar u^{(s)}_{\st}\}$ and $\bar v^{(s)}\vdash \{\bar v^{(s)}_{\so},\bar v^{(s)}_{\st}\}$ with $|\bar v^{(s)}_{\so}|=|\bar u^{(s)}_{\so}|$, $s=0,\dots,n-1$.

The functions $W(\bar v_{\so},\bar v_{\st}|\bar u_{\so},\bar u_{\st})$ do not depend on the eigenvalues $\lambda_j$, $j=0,\dots n$. As such, they do 
not depend on the model under consideration, they depend only on the $R$-matrix. 
\end{prop}
\proof
From lemma \ref{lem42}, we know that the scalar product is a sum of terms with functions $\alpha_s$ involving at most once each Bethe parameter $u_j^{(s)}$ or $v_k^{(s)}$. This can be realized as a sum over partitions as written in the expression \eqref{eq:scal1} with coefficients independent from the $\lambda$'s. Then, it remains to prove that the partitions obey the equalities $|\bar v^{(s)}_{\so}|=|\bar u^{(s)}_{\so}|$. 
For such a purpose, we
consider a composite model corresponding to a splitting  $T(u)=T^{[2]}(u)\, T^{[1]}(u)$ of the monodromy matrix, with $\alpha_s(u)=\alpha_s^{[1]}(u)\alpha_s^{[2]}(u)$. 

From the coproduct property for BVs \eqref{eq:copBB}, using \begin{equation*}
(\cdot)^{{\prime}}\otimes (\cdot)^{{\prime}}\,\circ\, \Delta = \Delta^{\rm op}\,\circ\,(\cdot)^{{\prime}}\,,
\end{equation*}
where $(\cdot)^{{\prime}}$ is the transposition anti-morphism 
defined by \r{a-mor},
we deduce 
\begin{equation}
\Delta(\CC(\bar v)) = \sum_{part} \Omega(\bar v_{\so}|\bar v_{\st}) \, \CC^{[1]}(\bar v_{\so})\, \CC^{[2]}(\bar v_{\st})\,\prod_{s=0}^{n-1} \alpha_s^{[1]}(\bar v^{(s)}_{\st})\,,
\label{eq:copCC}
\end{equation}
where $\Omega(\bar v_{\so}|\bar v_{\st})$ is defined as in \eqref{eq:defOmega}.
The sum in \eqref{eq:copCC} runs over partitions $\bar v^{(s)}\vdash \{\bar v^{(s)}_{\so},\bar v^{(s)}_{\st}\}$
and  $\CC^{[1]}(\bar v_{\so})$ and $\CC^{[2]}(\bar v_{\st})$ are dual BVs for the submodels based on 
$T^{[1]}(v)$ and $T^{[2]}(v)$ respectively. Then, the scalar product $\CC(\bar v)\,\BB(\bar u)$ takes the form
\begin{equation}\label{eq:scal2}
\begin{split}
S(\bar v|\bar u) = \Delta(S(\bar v|\bar u)) = \sum_{part} \Omega(\bar u_{\st}|\bar u_{\so}) \,\Omega(\bar v_{\so}|\bar v_{\st})\,S^{[1]}(\bar v_{\so}|\bar u_{\so})\, S^{[2]}(\bar v_{\st}|\bar u_{\st}) \,\prod_{s=0}^{n-1} \alpha_s^{[2]}(\bar u^{(s)}_{\so})\,\alpha_s^{[1]}(\bar v^{(s)}_{\st})\, .
\end{split}
\end{equation}
Since, from lemma \ref{lem:sym}, for the scalar products $S^{[1]}$ and $S^{[2]}$ to be non-zero, we need to have $|\bar v_{\so}^{(s)}|=|\bar u_{\so}^{(s)}|$ and $|\bar v_{\st}^{(s)}|=|\bar u_{\st}^{(s)}|$, we get the result.
\qed

\begin{prop}\label{prop:W}
The coefficients $W(\bar v_{\so},\bar v_{\st}|\bar u_{\so},\bar u_{\st})$ in $S(\bar v|\bar u) $ can be expressed as
\begin{equation}\label{eq:W}
 W(\bar v_{\so},\bar v_{\st}|\bar u_{\so},\bar u_{\st})=\ \Omega(\bar u_{\so}|\bar u_{\st}) \,\Omega(\bar v_{\st}|\bar v_{\so})\ Z(\bar v_{\so}|\bar u_{\so})\, \overline{Z}(\bar v_{\st}|\bar u_{\st})\,,
\end{equation}
where $\Omega$ is defined in \eqref{eq:defOmega}, and the highest coefficients $Z(\bar v_{\so}|\bar u_{\so})$ and $\overline{Z}(\bar v_{\st}|\bar u_{\st})$ are defined by
\begin{equation*}
Z(\bar v_{\so}|\bar u_{\so}) = W(\bar v_{\so},\varnothing|\bar u_{\so},\varnothing)\
\mb{and}
\overline Z(\bar v_{\st}|\bar u_{\st})= W(\varnothing,\bar v_{\st}|\varnothing,\bar u_{\st})\,,
\end{equation*}
with the normalisation $W(\varnothing,\varnothing|\varnothing,\varnothing)=1$.
\end{prop}
\proof
To prove this property, we use the fact that the coefficients $W(\bar v_{\so},\bar v_{\st}|\bar u_{\so},\bar u_{\st})$ do not depend on the model under consideration.
We fix two partitions $\bar u^{(s)}\vdash \{\bar u^{(s)}_{\rm i},\bar u^{(s)}_{\rm ii}\}$ and
$\bar v^{(s)}\vdash \{\bar v^{(s)}_{\rm i},\bar v^{(s)}_{\rm ii}\}$ such that $|\bar u^{(s)}_{\rm i}|=|\bar v^{(s)}_{\rm i}|$ and $|\bar u^{(s)}_{\rm ii}|=|\bar v^{(s)}_{\rm ii}|$, and consider a model\footnote{The existence of such model is shown in appendix \ref{sect:mod}.} where we have 
\begin{equation}\label{eq:alphas}
\alpha^{[1]}_s(z)=0\mb{if} z\in\bar v^{(s)}_{\rm ii} \mb{and} \alpha^{[2]}_s(z)=0\mb{if} z\in\bar u^{(s)}_{\rm i}\,.
\end{equation}
Considering expression \eqref{eq:scal1}, since $\alpha_s(z)=0$ when $z\in\bar v^{(s)}_{\rm ii}\cup \bar u^{(s)}_{\rm i}$, we obtain 
\begin{equation}\label{eq:scal-oneset}
S(\bar v|\bar u) =W(\bar v_{\rm i},\bar v_{\rm ii}|\bar u_{\rm i},\bar u_{\rm ii})\, \prod_{s=0}^{n-1}\alpha_s(\bar v_{\rm i}^{(s)})\,\alpha_s(\bar u_{\rm ii}^{(s)})\,. 
\end{equation}
On the other hand, we see that in \eqref{eq:scal2}, for the product $ \prod_{s=0}^{n-1}  \alpha_s^{[2]}(\bar u^{(s)}_{\so})\,\alpha_s^{[1]}(\bar v^{(s)}_{\st})$
to be non-zero, we need to have $\bar u^{(s)}_{\so}\subset\bar u^{(s)}_{\rm ii}$ and $\bar v^{(s)}_{\st}\subset\bar v^{(s)}_{\rm i}$. But because of the constraint on the cardinalities of the subsets  $|\bar u^{(s)}_{\so}|=|\bar v^{(s)}_{\so}|$ and $|\bar u^{(s)}_{\rm i}|=|\bar v^{(s)}_{\rm i}|$, it leads to
$\bar u^{(s)}_{\so}=\bar u^{(s)}_{\rm ii}$ and $\bar v^{(s)}_{\st}=\bar v^{(s)}_{\rm i}$, so that
\begin{equation}\label{eq:scal-oneset2}
\begin{split}
 S(\bar v|\bar u)=&  S^{[1]}(\bar v_{\rm ii}|\bar u_{\rm ii})\,S^{[2]}(\bar v_{\rm i}|\bar u_{\rm i})\, \Omega(\bar u_{\rm i}|\bar u_{{\rm ii}}) \Omega(\bar v_{\rm ii}|\bar v_{\rm i})
 \prod_{s=0}^{n-1}  \alpha_s^{[2]}(\bar u^{(s)}_{\rm ii})\,\alpha_s^{[1]}(\bar v^{(s)}_{\rm i})
\,.
\end{split}
\end{equation}
Now, we can use again relation \eqref{eq:scal1} to compute
 the scalar products $S^{[1]}(\bar v_{\rm ii}|\bar u_{\rm ii})$ and $S^{[2]}(\bar v_{\rm i}|\bar u_{\rm i})$. We detail the calculation of $S^{[1]}(\bar v_{\rm ii}|\bar u_{\rm ii})$, the other case being similar. We need to perform partitions $\bar u_{\rm ii}^{(s)}\vdash \{\bar u^{(s)}_{\rm iii},\bar u^{(s)}_{\rm iv}\}$ and
$\bar v_{\rm ii}^{(s)}\vdash \{\bar v^{(s)}_{\rm iii},\bar v^{(s)}_{\rm iv}\}$, which will make appear terms with a factor
$\prod_{s=0}^{n-1}\alpha^{[1]}_s(\bar v^{(s)}_{\rm iii})\,\alpha^{[1]}_s(\bar u^{(s)}_{\rm iv})$. However, because of  \eqref{eq:alphas}, only the partition corresponding to 
$\bar v^{(s)}_{\rm iii}=\varnothing$ will have a non-zero contribution. Since $|\bar v^{(s)}_{\rm iii}|=|\bar u^{(s)}_{\rm iii}|$, we get
\begin{equation}
\label{eq:scal-partiels}
\begin{split}
S^{[1]}(\bar v_{\rm ii}|\bar u_{\rm ii})&=W(\varnothing,\bar v_{\rm ii}|\varnothing,\bar u_{\rm ii})\,\prod_{s=0}^{n-1}\alpha^{[1]}_s(\bar u_{\rm ii}^{(s)})\,,\\
S^{[2]}(\bar v_{\rm i}|\bar u_{\rm i})&=W(\bar v_{\rm i},\varnothing|\bar u_{\rm i},\varnothing)\,\prod_{s=0}^{n-1}\alpha^{[2]}_s(\bar v_{\rm i}^{(s)})\,.
\end{split}
\end{equation}
Plugging these expressions in \eqref{eq:scal-oneset2} and comparing it with \eqref{eq:scal-oneset} gives relation \eqref{eq:W}.
\qed

\begin{coro}
The highest coefficient $Z(\bar v|\bar u)$ is invariant under any permutation of $\bar v^{(s)}$ and any permutation of $\bar u^{(s)}$, for any $s$.

It obey the symmetry relation
\begin{equation}\label{eq:symZ}
Z(\bar v|\bar u)=\overline Z(\bar u|\bar v)\,.
\end{equation}
This relation is sufficient to ensure the symmetry of the scalar product given in $(i)$  of lemma \ref{lem:sym}.
\end{coro}
\proof
From relation \eqref{eq:scal-partiels}, we deduce that there is a model where for instance
$$S^{[2]}(\bar v_{\rm i}|\bar u_{\rm i})=Z(\bar v_{\rm i}|\bar u_{\rm i})\,\prod_{s=0}^{n-1}\alpha^{[2]}_s(\bar v_{\rm i}^{(s)}).$$ The scalar product being invariant under the permutations, we deduce that in this model the highest coefficient is also invariant. Since the highest coefficient is independent from the choice of the model, we conclude it is always invariant.

Starting from symmetry relation $S(\bar u|\bar v)=S(\bar v|\bar u)$ and using the sum formula, we get
\begin{equation*}
\sum_{\rm part} W(\bar v_{\so},\bar v_{\st}|\bar u_{\so},\bar u_{\st})\, \prod_{s=0}^{n-1}\alpha_s(\bar v^{(s)}_{\so})\,\alpha_s(\bar u^{(s)}_{\st})
=
\sum_{\rm part} W(\bar u_{\so},\bar u_{\st}|\bar v_{\so},\bar v_{\st})\, \prod_{s=0}^{n-1}\alpha_s(\bar u^{(s)}_{\so})\,\alpha_s(\bar v^{(s)}_{\st})\,.
\end{equation*}
 Since $\alpha$'s are free functionals, we can project this relation on $\prod_{s=0}^{n-1}\alpha_s(\bar v^{(s)}_{\so})\,\alpha_s(\bar u^{(s)}_{\st})$ for a fixed but arbitrary partition. This leads to
\begin{equation} W(\bar v_{\so},\bar v_{\st}|\bar u_{\so},\bar u_{\st})=
W(\bar u_{\st},\bar u_{\so}|\bar v_{\st},\bar v_{\so}),\label{eq:symW}
\end{equation}
which is valid for any partition. Setting $\bar v_{\st}=\bar u_{\st}=\varnothing$ in this  relation, we get \eqref{eq:symZ}.
Note that from proposition \ref{prop:W}, one deduces that relation \eqref{eq:symZ} is sufficient to ensure  \eqref{eq:symW}  and implies the symmetry of the scalar product.
\qed

Gathering the results obtained in this section we get as a final formula for the scalar product
\begin{thm}\label{thm:sum-formula}
The scalar product $S(\bar v|\bar u)=\CC(\bar v)\,\BB(\bar u)$ obeys the following sum formula
\begin{equation}\label{eq:scal-fin}
S(\bar v|\bar u)=\sum_{part}\ \Omega(\bar u_{\so}|\bar u_{\st}) \,\Omega(\bar v_{\st}|\bar v_{\so})\ Z(\bar v_{\so}|\bar u_{\so})\, {Z}(\bar u_{\st}|\bar v_{\st})\,\prod_{s=0}^{n-1}\alpha_s(\bar v^{(s)}_{\so})\,\alpha_s(\bar u^{(s)}_{\st})\,,
\end{equation}
where the sum is taken over partitions $\bar u^{(s)}\vdash \{\bar u^{(s)}_{\so},\bar u^{(s)}_{\st}\}$ and $\bar v^{(s)}\vdash \{\bar v^{(s)}_{\so},\bar v^{(s)}_{\st}\}$ with $|\bar v^{(s)}_{\so}|=|\bar u^{(s)}_{\so}|$, $s=0,\dots,n-1$. 
$Z(\bar v|\bar u)$ is called the highest coefficient and does not depend on the choice of the model, i.e. it does not depend on the functions $\alpha_s$.
\end{thm}
Remark that the relation \eqref{eq:scal-fin} looks formally identical to the sum formula for $Y(\fg\fl_n)$ models, with however the restriction that the terms associated to $s=0$ have a different form, see definitions \eqref{eq:functions}.
This implies that when $\bar u^{(0)}=\bar v^{(0)}=\varnothing$, we get exactly the sum formula for the $Y(\fg\fl_{n})$ models, as expected.

\section{Recursion relations\label{sect:recZ}}

From the recurrence relations detailed in section \ref{sect:rec}, we can  deduce some recurrence relations obeyed by the scalar product. Their iterative action allows to express scalar products of $Y(\fo_{2n+1})$ models in terms of scalar products of $Y(\fg\fl_{n-\ell-1})\otimes Y(\fo_{2\ell+3})$ models. 
To lighten the presentation, we will use the  notation 
$\big(\bar v,z^{(\ell)}\big)$ introduced in \r{eq:def-z^ell}.
\begin{prop}\label{prop51}
For $\ell\geq0$, the scalar product obeys the following recurrence relation, with 
 $|\bar u^{(\ell)}|=|\bar v^{(\ell)}|+1$ and $|\bar u^{(s)}|=|\bar v^{(s)}|$ otherwise.
\begin{equation}\label{eq:recS-other}
\begin{split}
S(\bar v, z^{(\ell)}|\bar u) =\  \sum_{j=-n}^{\ell} \sum_{i=\ell+1}^{n} \sum_{\rm part} &
\frac{\left(\prod_{s=\ell+1}^{i-1}\alpha_s(\bar v^{(s)}_{\sth})\right)\left(\prod_{s=0}^{n-1}\alpha_s(\bar w^{(s)}_{\sth})\right)}{\prod_{s=\ell+1}^{n-1}\alpha_s(z)}\\
&\quad \times
\Phi_{i,j}(\bar w) \,\Psi^{(\ell)}_{j,i}(\bar v, z)
S(\bar v_{\st}|\bar w_{\st})\,,
\end{split}
\end{equation}
where the functions $\Phi_{i,j}(\bar w)$ and $\Psi^{(\ell)}_{j,i}(\bar v, z)$ are given in \eqref{eq:Phi-ij},  \eqref{eq:Psi-gen} and \eqref{eq:Psi-0}. 
The sum is on partitions $\bar w^{(s)}=\{\bar u^{(s)},z,z_s\}\vdash\{\bar w^{(s)}_{\so},\bar w^{(s)}_{\st},\bar w^{(s)}_{\sth}\}$ and
$\bar v^{(s)}\vdash\{\bar v^{(s)}_{\so},\bar v^{(s)}_{\st},\bar v^{(s)}_{\sth}\}$   
with cardinalities given in \eqref{eq:part-tot-TijB} and \eqref{eq:part-recCgen}. 
\end{prop}
\proof
Direct consequence of the recurrence relation \eqref{eq:recCCgen} and the action formula \eqref{eq:TijB}.
We used the relation
\begin{equation*}
\frac{\lambda_{\ell+1}(z)}{\lambda_n(z)}=\prod_{s=\ell+1}^{n-1}\alpha_s(z).
\end{equation*}\qed

Note that because of the scalar product $S(\bar v_{\st}|\bar w_{\st})$ in the rhs of the recurrence relations, we must also have 
$|\bar v_{\st}|=|\bar w_{\st}|$, which could lead to additional constraints on the partitions. We show it is not the case when $\ell=n-1$, the proof for the other values of $\ell$ being similar.
To compare the two sets $\bar v_{\st}$ and $\bar w_{\st}$, we use 
\begin{equation}\label{eq:part-action}
|\bar w^{(s)}_{\st}|=\begin{cases} |\bar w^{(s)}|-4=|\bar u^{(s)}|-2\,, \quad & j<-s \leq 0,\\ 
|\bar w^{(s)}|-3=|\bar u^{(s)}|-1\,, & |j|\leq s<n, \\ |\bar w^{(s)}|-2 =|\bar u^{(s)}|\,, &s<j ,\\ 0\,, &s=n, \end{cases}
\end{equation}
that is deduced from the cardinalities \eqref{eq:part-TijB}. Moreover, the lemma \ref{lem:sym}-$(iii)$ implies that
\begin{equation}\label{eq:part-scal}
|\bar v^{(s)}|=|\bar u^{(s)}|\,, \quad s\leq n-2\mb{and} |\bar v^{(n-1)}|=|\bar u^{(n-1)}|-1.
\end{equation}
It allows to
rewrite \eqref{eq:part-TijB} as
\begin{equation}\label{eq:part-action2}
\begin{split}
&\mb{For} s\leq n-2 : \qquad 
|\bar w^{(s)}_{\st}|=\begin{cases} |\bar v^{(s)}|-2, \quad &s<-j, \\  
|\bar v^{(s)}|-1, & |j|\leq s, \\  |\bar v^{(s)}|, & s<j. \end{cases}
\\[1ex]
&\mb{For} s= n-1 : \qquad 
|\bar w^{(n-1)}_{\st}|=\begin{cases} 
 |\bar v^{(n-1)}|-1,\quad &j=-n, \\
|\bar v^{(n-1)}|, & |j|\leq n-1.  \end{cases}
\end{split}
\end{equation}
On the other hand, from \eqref{eq:part-recCgen}, we deduce that 
\begin{equation}\label{eq:part-action3}
\begin{split}
&\mb{For} s\leq n-2 : \qquad 
|\bar v^{(s)}_{\st}|=\begin{cases}|\bar v^{(s)}|-2\,, \quad &j< -s\leq0\,, \\  |\bar v^{(s)}|-1\,, & |j|\leq s\,, \\ |\bar v^{(s)}|\,,\ & s<j \,.\end{cases}
\\[1ex]
&\mb{For} s= n-1 : \qquad 
|\bar v^{(n-1)}_{\st}|=\begin{cases} |\bar v^{(n-1)}|, & |j|\leq n-1,  \\ |\bar v^{(n-1)}|-1,\quad & j=n, \end{cases} \ 
\end{split}
\end{equation}
which shows that indeed we have $|\bar v_{\st}|=|\bar w_{\st}|$. Then,
 the only constraints on partitions in \eqref{eq:recS-other} are given by  \eqref{eq:part-tot-TijB} and \eqref{eq:part-recCgen}.

\subsection{Recursion for the highest coefficient}
There are different recursion relations for highest coefficients, depending on the color on which we perform the recursion.
It also depends on the way we compute the recursion, which amounts to say whether we deal with 
$Z(\bar v|\bar u)$ or $\overline{Z}(\bar v|\bar u)=Z(\bu|\bv)$.
\begin{prop}\label{prop52}
We remind the notation \eqref{eq:def-z^ell}.
The highest coefficient obeys the recurrence relation
\begin{equation}\label{eq:recZ-other}
Z(\bar u|\bar v, z^{(\ell)}) = \sum_{j=-n}^{\ell} \sum_{\rm part} 
\Phi_{\ell+1,j}(\bar x) \,\Psi^{(\ell)}_{j,\ell+1}(\bar v, z)\
Z(\bar x_{\st}|\bar v_{\st})\,,
\end{equation}
where 
\begin{equation}\label{x-set}
\bar x^{(s)}=\begin{cases} \bar u^{(s)}\ ,&\ 0\leq s\leq \ell, \\
\{\bar u^{(s)},z\}\ ,&\ \ell+1\leq s\leq n-1,
\end{cases}
\end{equation}
and the sum is on partitions $\bar x^{(s)}\vdash \{\bar x^{(s)}_{\st},\bar x^{(s)}_{\sth}\}$ and $\bar v^{(s)}\vdash\{\bar v^{(s)}_{\so},\bar v^{(s)}_{\st}\}$ with cardinalities
\begin{equation}\label{eq:part-rec-ell}
\begin{aligned}
&\mb{For} 0\leq s<\ell : \
&&|\bar v^{(s)}_{\so}|=
\begin{cases}2\,, &j< -s\,, \\  1\,, & -s\leq j\leq s\,, \\ 0\,,\ & s< j \,,\end{cases} 
\ & \big|\bar x^{(s)}_{\sth}\big|=\begin{cases} 2,\ &\ j< -s\,, \\  1,\ &\ -s\leq j\leq s\,, \\  0,\ &\ s< j\,,  \end{cases}
\\[1ex]
&\mb{For} \ell \leq s\leq n-1: \ 
&&|\bar v^{(s)}_{\so}|=
\begin{cases} 1,\  &\ j< -s\,, \\ 0,\ &\ -s\leq j \,,\end{cases} 
& \big|\bar x^{(s)}_{\sth}\big|=\begin{cases} 2,\ &\ j< -s\,, \\ 1,\ &\ -s\leq j\leq s\,,   \end{cases}
\end{aligned}
\end{equation}
with the convention $\bar v^{(n)}=\varnothing$ and $\bar x^{(n)}=\bar x^{(n)}_{\sth}=\{z\}$.
The function $\Phi_{\ell+1,j}(\bar x)$ is given by\footnote{Strictly speaking, there should be a factor $-\sigma_{-j}$ in $\Phi_{\ell+1,j}(\bar x)$ and 
a factor $\sigma_{j+1}$ in $\Psi^{(\ell)}_{j,\ell+1}(\bar v, z)$. However, in the product $\Phi_{\ell+1,j}(\bar x) \,\Psi^{(\ell)}_{j,\ell+1}(\bar v, z)$ 
these terms cancel since $-\sigma_{-j}\sigma_{j+1}=1$, so that we discarded them.} 
\begin{equation}\label{eq:Z-Phi-gen}
\Phi_{\ell+1,j}(\bar x) = g(z,\bar u^{(\ell)})\,h(\bar u^{(\ell+1)},z)\,
\Omega(\bx_{\st}|\bx_{\sth})\,,
\end{equation}
 while the function $\Psi^{(\ell)}_{j,\ell+1}(\bar v, z)$ reads 
 for $\ell>0$
\begin{equation}\label{eq:Z-Psi-ell}
\Psi^{(\ell)}_{j,\ell+1}(\bar v, z) =  \frac{\Omega(\bv_{\so}|\bv_{\st})}{g(z,\bar v^{(\ell-1)}_{\st})\,h(z,\bar v^{(\ell)})\,h(\bar v^{(\ell)}_{\st},z)\,g(\bar v^{(\ell+1)},z)}\,,
\end{equation}
and for $\ell=0$
\begin{equation}\label{eq:Z-Psi-0}
\Psi^{(0)}_{j,1}(\bar v, z)= \frac{g(z_0,\bar v^{(0)}_{\so})\,\Omega(\bv_{\so}|\bv_{\st})}{\ff(z_0,\bar v^{(0)})\,g(\bar v^{(1)},z)}
\,.
\end{equation}
\end{prop}
\proof
We consider expressions modulo terms that contain at least a $\alpha_s( v^{(s)}_k)$ or a $\alpha_s(z)$ or a $\alpha_s(z_s)$ term. We note $\simeq$ this equivalence relation. The sum formula \eqref{eq:scal-fin} leads to
\begin{equation}\label{eq:S-HCu2}
S(\bar v,z^{(\ell)}|\bar u)\simeq  Z(\bar u|\bar v,z^{(\ell)})\,\prod_{s=0}^{n-1}\alpha_s(\bar u^{(s)})\,.
\end{equation}

Starting from  the recurrence relation \eqref{eq:recS-other} and using \eqref{eq:scal-fin} for the scalar product $S(\bar v_{\st}|\bar w_{\st})$, the scalar product
$S(\bar v, z^{(\ell)}|\bar u)$ also reads 
\begin{equation*}%\label{}
\begin{split}
S(\bar v, z^{(\ell)}|\bar u) &=\  \sum_{j=-n}^{\ell} \sum_{i=\ell+1}^{n} \sum_{\rm part} 
\frac{\left( \prod_{s=\ell+1}^{i-1}\alpha_s(\bar v^{(s)}_{\sth})\right)\left(\prod_{s=0}^{n-1}\alpha_s(\bar w^{(s)}_{\sth})\right)}{\prod_{s=\ell+1}^{n-1}\alpha_s(z)}\times\\
&\qquad\times
\Phi_{i,j}(\bar w) \,\Psi^{(\ell)}_{j,i}(\bar v, z)
S(\bar v_{\st}|\bar w_{\st})
\\
&\simeq\  \sum_{j=-n}^{\ell} \sum_{i=\ell+1}^{n} \sum_{\rm part} 
\frac{\left(\prod_{s=\ell+1}^{i-1}\alpha_s(\bar v^{(s)}_{\sth})\right)\left(\prod_{s=0}^{n-1}\alpha_s(\bar w^{(s)}_{\sth})\right)}{\prod_{s=\ell+1}^{n-1}\alpha_s(z)}\times\\
&\qquad\times\Phi_{i,j}(\bar w) \,\Psi^{(\ell)}_{j,i}(\bar v, z)\prod_{s=0}^{n-1}\alpha_s(\bar w^{(s)}_{\st})\
Z(\bar w_{\st}|\bar v_{\st})\,.
\end{split}
\end{equation*}
To compare the two expressions, we need to select the partitions such that 
\begin{subequations}
\begin{eqnarray}
&&|\bar v^{(s)}_{\sth}|=0\mb{for}\ell+1\leq s\leq {i-1}\,,\label{eq:cond-v}\\
&&\{\bar w^{(s)}_{\st},\bar w^{(s)}_{\sth}\}=\bar u^{(s)} \mb{for} 0\leq s\leq \ell\,,\label{eq:cond-w-1} \\
&&\{\bar w^{(s)}_{\st},\bar w^{(s)}_{\sth}\}=\{\bar u^{(s)},z\} \mb{for} \ell+1\leq s\leq n-1\label{eq:cond-w-2}\,.
\end{eqnarray}
\end{subequations}
Looking at the cardinalities \eqref{eq:part-recCgen} for $\bar v$, one sees that $|\bar v^{(s)}_{\sth}|=0$ most of the time. The only cases when we have $|\bar v^{(s)}_{\sth}|=1$ correspond to $\ell<s<i$. But this cannot occur because the sum over $i$ runs from $\ell+1$ to $n$. In other words, the condition \eqref{eq:cond-v} is always obeyed.
Regarding the parameters $\bar w$, conditions \eqref{eq:cond-w-1} and \eqref{eq:cond-w-2} imply that $\bar w^{(s)}_{\so}=\{z,z_s\}$ for $0\leq s\leq \ell$ and 
$\bar w^{(s)}_{\so}=\{z_s\}$ for $\ell+1\leq s\leq n-1$. Then, we can replace the sum over partitions of $\bar w^{(s)}$ by a sum over partitions of $\bar x^{(s)}=\bar u^{(s)}$ for $0\leq s\leq \ell$ and $\bar x^{(s)}=\{\bar u^{(s)},z\}$ for $\ell+1\leq s\leq n-1$.
Moreover, looking at the cardinalities \eqref{eq:part-tot-TijB} for $\bar w^{(s)}_{\so}$, we see that we must take $i=\ell+1$ 
to have cardinalities compatible with the sets $\bar w^{(s)}_{\so}$ given above. 

From the subsets $\sw{s}_{\so}$ and $\sx{s}=\{\sw{s}_{\st},\sw{s}_{\sth}\}$
\begin{equation}\label{set1}
\sw{s}_{\so}=\begin{cases} \{z,z_s\},&0\leq s\leq\ell,\\ 
\{z_s\},&\ell+1\leq s\leq n-1,\end{cases}\qquad
\sx{s}=\begin{cases}\su{s},&0\leq s\leq\ell,\\ 
\{\su{s},z\},&\ell+1\leq s\leq n-1,\end{cases}
\end{equation}
described above 
and the boundary subsets $\sw{n}_{\so}=\{z_n\}$ and $\sx{n}=\sw{n}_{\sth}=
\{z\}$ one can calculate 
\begin{equation}\label{cal1}
\Omega(\bw_{\so}|\bw_{\st})\,\Omega(\bw_{\so}|\bw_{\sth})=
\Omega(\bw_{\so}|\bx)=\kappa\ \frac{h(z,\su{0})}{g(z_1,\su{0})}\ 
g(z,\su{\ell})\,h(\su{\ell+1},z)
\end{equation}
which proves the equality \r{eq:Z-Phi-gen} from \r{eq:Phi-ij}. The equalities 
\r{eq:Z-Psi-ell} and \r{eq:Z-Psi-0} follows from the definitions 
of the functions $\Psi^{(\ell)}_{j,i}(\bv,z)$ 
\r{eq:Psi-gen} and \r{eq:Psi-0} since $\sv{s}_{\sth}=\varnothing$ 
for all $s=0,1,\ldots,n-1$.

Altogether, we get
\begin{equation*}%\label{}
\begin{split}
&S(\bar v, z^{(\ell)}|\bar u) \simeq\  \prod_{s=0}^{n-1}\alpha_s(\bar u^{(s)})
\sum_{j=-n}^{\ell} \sum_{\rm part} 
\Phi_{\ell+1,j}(\bar x) \,\Psi^{(\ell)}_{j,\ell+1}(\bar v, z)\
Z(\bar x_{\st}|\bar v_{\st})\,,
\end{split}
\end{equation*}
which leads to the relation \eqref{eq:recZ-other}. 
\qed

\begin{prop}\label{prop53}
The highest coefficient also obeys the following recurrence relation
\begin{equation}\label{eq:recZ-other2}
Z(\bar v, z^{(\ell)}|\bar u) = \sum_{i=\ell+1}^{n} \sum_{\rm part} 
\Phi_{i,\ell}(\bar w) \,\Psi^{(\ell)}_{\ell,i}(\bar v, z)\,Z(\bar v_{\st}|\bar w_{\st})
\,,
\end{equation}
with for $\ell>0$
\begin{equation}\label{eq:Phi-iell}
\Phi_{i,\ell}(\bar w)= \frac{g(z_1,\bar u^{(0)})}{h(z,\bar u^{(0)})}\ 
h(z,\bar u^{(\ell-1)})\,g(\bar u^{(\ell)},z)\ 
\Omega(\bw_{\so}|\bw_{\st})\,,
\end{equation}
\begin{equation}\label{eq:Psi-elli}
\Psi^{(\ell)}_{\ell,i}(\bar v, z)= \frac{\Omega(\bv_{\st}|\bv_{\sth})}
{g(z,\bar v^{(\ell-1)})\,h(z,\bar v^{(\ell)})\,h(\bar v^{(\ell)},z)\,g(\bar v^{(\ell+1)}_{\st},z)}\,,
\end{equation}
while for $\ell=0$
\begin{equation}\label{eq:Phi-i0}
\Phi_{i,0}(\bar w)=\ -\ g(z,\bar u^{(0)})\ \Omega(\bw_{\so}|\bw_{\st})\,,
\end{equation}
\begin{equation}\label{eq:Psi-0i}
\Psi^{(0)}_{0,i}(\bar v, z) =
\frac{\Omega(\bv_{\st}|\bv_{\sth})}{\ff(z_0,\bar v^{(0)})g(\bar v^{(1)}_{\st},z)}\,.
\end{equation}
The sum is over partitions $\bar v^{(s)}\vdash\{\bar v^{(s)}_{\st},\bar v^{(s)}_{\sth}\}$ and $\bar w^{(s)}\vdash\{\bar w^{(s)}_{\so},\bar w^{(s)}_{\st}\}$ with 
\begin{equation}\label{w-set}
\bar w^{(s)}=\{\bar u^{(s)},z,z_s\}\mb{for}0\leq s<\ell\,,\qquad \bar w^{(s)}=\{\bar u^{(s)},z_s\}\mb{for}\ell\leq s\leq n-1\,. 
\end{equation}
The cardinalities are given by
\begin{equation}\label{eq:part-Psi-elli}
\mbox{for }\ 0\leq s\leq\ell : \quad 
|\bar v^{(s)}_{\sth}|=0 ; \quad\quad
\mbox{for }\ \ell <s\leq n-1: \quad 
|\bar v^{(s)}_{\sth}|=
\begin{cases} 1, & s< i, \\ 0,  & i\leq s, \end{cases}
\end{equation} 
and (for $i\geq\ell+1$)
\begin{equation}\label{eq:part-Phi-iell}
\big|\bar w^{(s)}_{\so}\big|=\begin{cases} 2,\ &\ s<i \,, \\  1,\ &\  i\leq s \,, \end{cases}
\quad  \quad \bar w^{(s)}_{\sth}=\begin{cases} \varnothing,\ &\ s< \ell \,, \\  \{z\},\ &\ \ell\leq s\,.  \end{cases}
\end{equation}
\end{prop}
\proof
From \eqref{eq:scal-fin},  we  have
\begin{equation*}%\label{}
S(\bar v, z^{(\ell)}|\bar u)\sim Z(\bar v, z^{(\ell)}|\bar u)\,\alpha_{\ell}(z)\,\prod_{s=0}^{n-1}\alpha_s(\bar v^{(s)})\,,
\end{equation*}
where $\sim$ stands for expressions modulo terms that contain at least  one factor 
$\alpha_s(u^{(s)}_j)$ for some $j$ and some $s$, or at least one factor $\alpha_s(z)$ for some $s\neq\ell$, or at least one 
factor\footnote{Factors $\alpha_s(z_s)$ do not appear in \eqref{eq:scal1}, so that this requirement is useless here. However, this exclusion is needed in the following.}  $\alpha_s(z_s)$ for some $s$. 
Starting from  the recurrence relation \eqref{eq:recS-other} and using \eqref{eq:scal-fin} for the scalar product $S(\bar v_{\st}|\bar w_{\st})$,  we get
\begin{equation*}%\label{}
\begin{split}
S(\bar v, z^{(\ell)}|\bar u) &=\  \sum_{j=-n}^{\ell} \sum_{i=\ell+1}^{n} \sum_{\rm 
part} 
\frac{\left(\prod_{s=\ell+1}^{i-1}\alpha_s(\bar v^{(s)}_{\sth})\right)\left(\prod_{s=0}^{n-1}\alpha_s(\bar w^{(s)}_{\sth})\right)}{\prod_{s=\ell+1}^{n-1}\alpha_s(z)}\times\\
&\qquad\times\Phi_{ij}(\bar w) \,\Psi^{(\ell)}_{ji}(\bar v, z)
S(\bar v_{\st}|\bar w_{\st})
\\
&\sim\  \sum_{j=-n}^{\ell} \sum_{i=\ell+1}^{n} \sum_{\rm part} 
\frac{\left(\prod_{s=\ell+1}^{i-1}\alpha_s(\bar v^{(s)}_{\sth})\right)\left(\prod_{s=0}^{n-1}\alpha_s(\bar w^{(s)}_{\sth})\right)}{\prod_{s=\ell+1}^{n-1}\alpha_s(z)}\times\\
&\qquad\times\Phi_{ij}(\bar w) \,\Psi^{(\ell)}_{ji}(\bar v, z)\prod_{s=0}^{n-1}\alpha_s(\bar v^{(s)}_{\st})\
Z(\bar v_{\st}|\bar w_{\st})\,.
\end{split}
\end{equation*}
This implies that we must have $\bar w^{(s)}_{\sth}=\varnothing$ for $0\leq s<\ell$ and $\bar w^{(s)}_{\sth}=\{z\}$ for $\ell\leq s\leq n-1$.
Looking at the cardinalities \eqref{eq:part-tot-TijB}, it implies that we must have $j\geq \ell$. Since the sum over $j$ runs up to $\ell$, it implies that $j=\ell$, so that 
\begin{equation}%\label{}
\begin{split}
S(\bar v, z^{(\ell)}|\bar u) &\sim\   \sum_{i=\ell+1}^{n} \sum_{part} 
\Phi_{i\ell}(\bar w) \,\Psi^{(\ell)}_{\ell i}(\bar v, z)\,Z(\bar v_{\st}|\bar w_{\st})\,\alpha_\ell(z)
\prod_{s=0}^{n-1}\alpha_s(\bar v^{(s)})
\,.
\end{split}
\end{equation}
From \eqref{eq:Phi-ij} and \eqref{eq:part-tot-TijB}, we obtain \eqref{eq:Phi-iell} and the cardinalities \eqref{eq:part-Phi-iell}. Looking at the $\bar v$ part, since $j=\ell$, one sees that $|\bar v^{(s)}_{\so}|=0$ for any $s$, and we get a partition $\bar v^{(s)}\vdash\{\bar v^{(s)}_{\st}, \bar v^{(s)}_{\sth}\}$ with cardinalities \eqref{eq:part-Psi-elli}, deduced from \eqref{eq:part-recCgen} for the specific values of $i$ and $j$. Then, from \eqref{eq:Psi-gen} and \eqref{eq:Psi-0} we obtain \eqref{eq:Psi-elli} and \eqref{eq:Psi-0i}.
\qed

\subsection{Reduction to $Y(\fg\fl_{n})$ models\label{sec:gln}}
It has been shown in \cite{LPR2024}, that when $\ell>0$ and there is no Bethe parameters of color 0, the recurrence relations for Bethe vectors in $Y(\fo_{2n+1})$ models reproduce the recurrence relations for Bethe vectors in $Y(\fg\fl_{n})$ models. Since the starting point (the vacuum state) is identical, it implies that the Bethe vectors in $Y(\fo_{2n+1})$ models, when they have no Bethe parameters of color 0, are identical to the ones of $Y(\fg\fl_{n})$ models. Indeed, such Bethe vectors are built using solely the entries $T_{ij}(z)$, $1\leq i\leq j\leq n$ of the monodromy matrix. It is easy to see from \eqref{comtt} that these entries obey the commutation relations of $Y(\fg\fl_{n})$, showing the embedding $Y(\fg\fl_{n})\subset Y(\fo_{2n+1})$. 

Thus we expect this relation to hold also at the level of scalar products and highest coefficients. In other words,
for $\ell>0$ the recurrence relations  \r{eq:recZ-other}
and \r{eq:recZ-other2} may be compared with the recurrence relations 
for the highest coefficients obtained in the papers 
\cite{HLPRS2017a,LP2021}. In the first paper only the case 
$\ell=n-1$ was considered, while in the second paper the general 
case of $1\leq\ell\leq n-1$ was investigated in the case of 
$U_q(\fg\fl_n)$ invariant models. These results can be easily translated 
to the case of $\fg\fl_n$ invariant models and compared with 
the results obtained above and specialized to the  $\fg\fl_n$ case.

To perform this specialization
we consider the above relations for the highest coefficient in the particular case $\bar v^{(0)}=\bar u^{(0)}=\varnothing$. In that case $\bar v^{(0)}_{\st}=\varnothing$, which implies that $\bar w^{(0)}_{\st}=\varnothing$
and the  relations are valid for highest coefficients of $Y(\fg\fl_{n})$ models. 
We detail them below. We will also show that they generalize some formulas already obtained in the context of $Y(\fg\fl_{n})$ models.

\begin{prop}\label{prop54}
For $Y(\fg\fl_n)$ models, the highest coefficient obeys the recurrence relation for $\ell>0$
\begin{equation}\label{eq:recZgln-2-ell}
Z(\bar u|\bar v, z^{(\ell)}) = \sum_{j=1}^{\ell} \sum_{\rm part} 
\Phi_{\ell+1,j}(\bar x) \,\Psi^{(\ell)}_{j,\ell+1}(\bar v, z)\
Z(\bar x_{\st}|\bar v_{\st})\,,
\end{equation}
where
\begin{equation}\label{eq:phi-ell-gln1}
\Phi_{\ell+1,j}(\bar x) = g(z,\bar u^{(\ell)})\,h(\bar u^{(\ell+1)},z)
\prod_{s=j}^{n} \gamma(\bar x^{(s)}_{\st},\bar x^{(s)}_{\sth})\,
\frac{\prod_{s=j}^n h(\bx^{(s)}_{\sth},\bx^{(s-1)}_{\st})}
{\prod_{s=j}^{n-2} g(\bx^{(s+1)}_{\st},\bx^{(s)}_{\sth})}\,,
\end{equation}
\begin{equation}\label{eq:Psi-ell-gln1}
\begin{split}
\Psi^{(\ell)}_{j,\ell+1}(\bar v, z) &= \Big(g(z,\bar v^{(\ell-1)}_{\st})\,
h(z,\bar v^{(\ell)})\,h(\bar v^{(\ell)},z)\,g(\bar v^{(\ell+1)},z)\Big)^{-1}
\times \\
&\qquad\times\ \prod_{s=j}^{\ell-1} \gamma(\bar v^{(s)}_{\so},\bar v^{(s)}_{\st})\,
\frac{h(\bar v^{(s+1)}_{\st},\bar v^{(s)}_{\so})}{g(\bar v^{(s+1)}_{\so},\bar v^{(s)}_{\st})}\,.
\end{split}
\end{equation}
The summation is over partitions of $\bar x^{(s)}\vdash \{\bar x^{(s)}_{\st},\bar x^{(s)}_{\sth}\}$ and $\bar v^{(s)}\vdash \{\bar v^{(s)}_{\so},\bar v^{(s)}_{\st}\}$ with cardinalities given in \eqref{eq:part-rec-ell}.
We recall that $\bar x^{(s)}=\bar u^{(s)}$ for $s\leq \ell$ and $\bar x^{(s)}=\{\bar u^{(s)},z\}$ for $s>\ell$.
\end{prop}
\proof
We consider the recurrence relation of the highest coefficient, equation \eqref{eq:recZ-other}, in the particular case $\bar x^{(0)}=\bar v^{(0)}=\bar u^{(0)}=\varnothing$, which implies that $\bar x^{(0)}_{\st}=\bar x^{(0)}_{\sth}=\varnothing$.  Since $\bar x^{(0)}_{\sth}=\varnothing$, the cardinalities 
 \eqref{eq:part-rec-ell} imply that we must take $j>0$, and that there is no partition of $\bar x^{(s)}$ when $s<j$. 
 Moreover, since $j\leq \ell$ and $\bar x^{(s)}=\bar u^{(s)}$ when $s\leq \ell$, in \eqref{eq:Z-Phi-gen} we can split the product on $s$ in two parts. It leads to \eqref{eq:phi-ell-gln1}. To get \eqref{eq:Psi-ell-gln1}, one notices that $|\bar v^{(s)}_{\so}|=0$ if $s<j$ or $s\geq \ell$.
\qed

\begin{prop}\label{prop55}
For $Y(\fg\fl_n)$ models and $\ell>0$, the highest coefficient also obeys the recurrence relation
\begin{equation}\label{eq:recZgln-2}
Z(\bar u|\bar v, z^{(\ell)}) = \sum_{i=\ell+1}^{n} \sum_{\rm part} 
\Phi_{i,\ell}(\bar x) \,\Psi^{(\ell)}_{\ell,i}(\bar v, z)\,Z(\bar v_{\st}|\bar x_{\st})\,,
\end{equation}
where
\begin{equation}\label{eq:phi-ell-gln2}
\Phi_{i,\ell}(\bar x)=h(z,\bar u^{(\ell-1)})g(\bar u^{(\ell)},z)
\prod_{s=1}^{i-1} \gamma(\bar x^{(s)}_{\so},\bar x^{(s)}_{\st})
\frac{\prod_{s=0}^{i-1}h(\bar x^{(s+1)}_{\st},\bar x^{(s)}_{\so})}
{\prod_{s=1}^{i-2}g(\bar x^{(s+1)}_{\so},\bar x^{(s)}_{\st})}
\end{equation}
\begin{equation}\label{eq:psi-ell-gln2}
\begin{split}
\Psi^{(\ell)}_{\ell,i}(\bar v, z) &= 
\Big(g(z,\bar v^{(\ell-1)})\,h(z,\bar v^{(\ell)})\,h(\bar v^{(\ell)},z)\,
g(\bar v^{(\ell+1)}_{\st},z)\Big)^{-1}\times\\
&\qquad\times
\prod_{s=\ell+1}^{i-1} \gamma_s(\bar v^{(s)}_{\st},\bar v^{(s)}_{\sth})\,
\frac{h(\bar v^{(s)}_{\sth},\bar v^{(s-1)}_{\st})}
{g(\bar v^{(s+1)}_{\st},\bar v^{(s)}_{\sth})} \,.
\end{split}
\end{equation}
The sum runs over partitions 
$\bar x^{(s)}\vdash\{\bar x^{(s)}_{\so},\bar x^{(s)}_{\st},\bar x^{(s)}_{\sth}\}$ and 
$\bar v^{(s)}\vdash\{\bar v^{(s)}_{\so},\bar v^{(s)}_{\st},\bar v^{(s)}_{\sth}\}$, where $\bar x^{(s)}=\bar u^{(s)}$  for $s<\ell$ and $\bar x^{(s)}=\{\bar u^{(s)},z\}$ for $s\geq\ell$. 
The cardinalities are given by
\begin{equation}
\begin{split}
&\big|\bar x^{(s)}_{\so}\big|=\begin{cases} 1,\ &\ 0<s<i \,, \\  0,\ &\ \ell+1\leq i\leq s\,, \end{cases}\qquad
\bar x^{(s)}_{\sth}=\begin{cases} \varnothing,\ &\ 0<s<\ell \,, \\  \{z\},\ &\ \ell\leq s\,, \end{cases}
\\
&\big|\bar v^{(s)}_{\so}\big|=0,\,\ \ \forall s, \qquad\qquad\qquad\qquad
\big|\bar v^{(s)}_{\sth}\big|=\begin{cases} 1,\ &\ \ell<s<i \,, \\  0,\ &\ \textrm{otherwise} \,. \end{cases}
\end{split}
\end{equation}
\end{prop}
\proof
We consider the recurrence relation of the highest coefficient, equation \eqref{eq:recZ-other2}, in the particular case $\bar v^{(0)}=\bar u^{(0)}=\varnothing$. In that case, $\bar w^{(0)}=\{z,z_0\}$ with $|\bar w^{(0)}_{\so}|=2$, so that there is no partition for $\bar w^{(0)}$. Then, as a first step, we get 
\begin{equation}%\label{}
\begin{split}
\Phi_{i\ell}(\bar w)&=\ h(z,\bar u^{(\ell-1)})g(\bar u^{(\ell)},z)h(\bar w^{(1)}_{\st},z)h(\bar w^{(1)}_{\st},z_0)
\times\\
&\quad\times
\prod_{s=1}^{n-1} \gamma_s(\bar w^{(s)}_{\so},\bar w^{(s)}_{\st})\,
\frac{h(\bar w^{(s+1)}_{\st},\bar w^{(s)}_{\so})}{g(\bar w^{(s+1)}_{\so},\bar w^{(s)}_{\st})}
\,.
\end{split}
\end{equation}
The factor $h(\bar w^{(1)}_{\st},z_0)$ implies that $z_1\not\in \bar w^{(1)}_{\st}$ so that $z_1\in \bar w^{(1)}_{\so}$. 
Then due to the factor $h(\bar w^{(s+1)}_{\st},\bar w^{(s)}_{\so})$, we get $z_s\in \bar w^{(s)}_{\so}$. Hence, we can replace the sum on partitions
of $\bar w^{(s)}=\{\bar u^{(s)},z,z_s\}$ by a partition on $\bar x^{(s)}$ with  $\bar x^{(s)}=\bar u^{(s)}$  for $s<\ell$ and $\bar x^{(s)}=\{\bar u^{(s)},z\}$ for $s\geq\ell$. The connexion between $\bar w^{(s)}$ and $\bar x^{(s)}$ is given by $\bar w^{(s)}_{\so}=\{\bar x^{(s)}_{\so},z_s\}$, $\bar w^{(s)}_{\st}=\bar x^{(s)}_{\st}$ and $\bar w^{(s)}_{\sth}=\bar x^{(s)}_{\sth}$. 
It leads to  \eqref{eq:phi-ell-gln2}. The cardinalities for $\bar x$ are deduced from those of $\bar w$, see \eqref{eq:part-Phi-iell}. It allows to reduce the product on $s$ from $1\leq s\leq n-1$ to $1\leq s\leq i-1$.

In the same way, the cardinalities on $\bar v$ are deduced from \eqref{eq:part-Psi-elli} and allows to reduce the product on $s$.
\qed

The rational version of the BV scalar product computed in the paper 
\cite{LP2021} reads 
\begin{equation}\label{c1}
\begin{split}
\tilde{S}(\bv|\bu)&=\tilde{\CC}(\bv)\,\tilde{\BB}(\bu)=
\sum_{\rm part} \tilde{Z}(\bv_{\so}|\bu_{\so})\ \tilde{Z}(\bu_{\st}|\bv_{\st})\
\prod_{s=1}^{n-1}\alpha_s(\sv{s}_{\st})^{-1}\,\alpha_s(\su{s}_{\so})^{-1}\times\\
&\times \prod_{s=1}^{n-1}
\Big(f(\sv{s}_{\st},\sv{s}_{\so})\,f(\su{s}_{\so},\su{s}_{\st})\Big)
\prod_{s=1}^{n-2}
\Big(f(\sv{s+1}_{\st},\sv{s}_{\so})\,f(\su{s+1}_{\so},\su{s}_{\st})\Big)^{-1}.
\end{split}
\end{equation}

In order to compare this scalar product with the scalar product \r{eq:scal-fin}
one has to renormalize the Bethe and dual Bethe vectors as follows 
\begin{equation}\label{c2}
\begin{split}
\BB(\bu)&=\prod_{s=1}^{n-1}\alpha_s(\su{s})
\prod_{s=1}^{n-2}h(\su{s+1},\su{s})
\prod_{s=1}^{n-1}h(\su{s},\su{s})^{-1}\ \tilde\BB(\bu)\,,\\
\CC(\bv)&=\prod_{s=1}^{n-1}\alpha_s(\sv{s})
\prod_{s=1}^{n-2}h(\sv{s+1},\sv{s})
\prod_{s=1}^{n-1}h(\sv{s},\sv{s})^{-1}\ \tilde\CC(\bv)\,.
\end{split}
\end{equation}
Comparing now the scalar product \r{eq:scal-fin} for these Bethe vectors 
with the scalar product \r{c1} one gets the relation between 
the highest coefficients $Z(\bv|\bu)$ and $\tilde{Z}(\bv|\bu)$
\begin{equation}\label{c3}
Z(\bu|\bv)=\tilde{Z}(\bu|\bv)\ \prod_{s=1}^{n-1}
\frac{h(\su{s+1},\su{s})\,h(\sv{s+1},\sv{s})}
{h(\su{s},\su{s})\,h(\sv{s},\sv{s})}
\end{equation}
which allows to compare the recurrence relations obtained in \cite{LP2021}
and specialized to the rational case 
 with the relations \r{eq:recZgln-2-ell} and \r{eq:recZgln-2} above. 

After renaming 
the sets of  Bethe parameters,  
one of the recurrence relation for the highest coefficient found 
in the paper \cite{LP2021}
  can be written in the following form 
  \begin{equation}\label{c4}
\begin{split}
\tilde{Z}(\bu|\bv,z^{(\ell)})&=
\frac{1}{f(z,\sv{\ell-1})\,f(\sv{\ell+1},z)}
\sum_{j=1}^\ell\sum_{\rm part}
\prod_{s=j}^{\ell-1}
\frac{g(\sv{s+1}_{\so},\sv{s}_{\so})\,f(\sv{s}_{\so},\sv{s}_{\st})}
{f(\sv{s}_{\so},\sv{s-1})}
\times\\
&\times 
 f(z,\su{\ell}) \prod_{s=j}^{n-1}
\frac{g(\sx{s+1}_{\sth},\sx{s}_{\sth})\,f(\sx{s}_{\st},\sx{s}_{\sth})}
{f(\sx{s+1},\sx{s}_{\sth})}\times\\
&\times \tilde{Z}\Big(\{\sx{s}\}_1^{j-1},\{\sx{s}_{\st}\}_j^{n-1}|
\{\sv{s}\}_1^{j-1},\{\sv{s}_{\st}\}_j^{\ell-1},\{\sv{s}\}_\ell^{n-1}\Big)
\end{split}
\end{equation}
where in \r{c4} $\sv{\ell}_{\so}\equiv \{z\}$, 
$\sv{\ell}_{\st}\equiv \sv{\ell}$ and the summation  goes over 
partitions of the sets described in proposition~\ref{prop54}.  
The recurrence relation \r{c4} for $\ell=n-1$ coincides  with 
the relation (4.26) of \cite{HLPRS2017a}.

Multiplying now both sides of the recurrence relation \r{c4} by the 
product 
\begin{equation}\label{c5}
\frac{h(z,\sv{\ell-1})\,h(\sv{\ell+1},z)}{h(z,\sv{\ell})\,h(\sv{\ell},z)}
\prod_{s=1}^{n-1}\frac{h(\sv{s+1},\sv{s})}{h(\sv{s},\sv{s})}
\prod_{s=1}^{n-1}\frac{h(\su{s+1},\su{s})}{h(\su{s},\su{s})}
\end{equation}
and using the relation  
\begin{equation}\label{c6}
 \prod_{s=1}^{n-1}\frac{h(\su{s+1},\su{s})}{h(\su{s},\su{s})}=
 \frac{h(\su{\ell+1},z)}{h(z,\su{\ell})}\ 
 \prod_{s=1}^{n-1}\frac{h(\sx{s+1},\sx{s})}{h(\sx{s},\sx{s})}
\end{equation}
which follows from definition of the sets 
$\sx{s}$, one gets the recurrence relations \r{eq:recZgln-2-ell} with the functions 
$\Phi_{\ell+1,j}(\bw)$ and $\Psi^{(\ell)}_{j,\ell+1}(\bv,z)$ given 
by equalities \r{eq:phi-ell-gln1} and \r{eq:Psi-ell-gln1}
respectively. 

Analogously, renaming the sets of  Bethe parameters 
in the second recurrence relation for the highest coefficient in 
\cite{LP2021}, one can present it in the form  
\begin{equation}\label{c7}
\begin{split}
\tilde{Z}(\bv,z^{(\ell)}|\bu)&=
\frac{1}{f(z,\sv{\ell-1})\,f(\sv{\ell+1},z)}
\sum_{i=\ell+1}^n\sum_{\rm part}
\prod_{s=\ell+1}^{i-1}
\frac{g(\sv{s}_{\sth},\sv{s-1}_{\sth})\,f(\sv{s}_{\st},\sv{s}_{\sth})}
{f(\sv{s+1},\sv{s}_{\sth})}
\times\\
&\qquad\times 
 f(\su{\ell},z) \prod_{s=1}^{i-1}
\frac{g(\sx{s}_{\so},\sx{s-1}_{\so})\,f(\sx{s}_{\so},\sx{s}_{\st})}
{f(\sx{s}_{\so},\sx{s-1})}\times\\
&\qquad\times \tilde{Z}\Big(
\{\sv{s}\}_1^{\ell},\{\sv{s}_{\st}\}_{\ell+1}^{i-1},\{\sv{s}\}_i^{n-1}|
\{\sx{s}\}_1^{i-1},\{\sx{s}_{\st}\}_i^{n-1}
\Big)
\end{split}
\end{equation}
where  again in \r{c7} $\sv{\ell}_{\so}\equiv \{z\}$, 
$\sv{\ell}_{\st}\equiv \sv{\ell}$ and the summation  goes over 
partitions of the sets described in proposition~\ref{prop55}. 
Multiplying both sides of the recurrence equality \r{c7}
by the product \r{c5} and using the relation 
\begin{equation}\label{c8}
 \prod_{s=1}^{n-1}\frac{h(\su{s+1},\su{s})}{h(\su{s},\su{s})}=
 \frac{h(z,\su{\ell-1})}{h(\su{\ell},z)}\ 
 \prod_{s=1}^{n-1}\frac{h(\sx{s+1},\sx{s})}{h(\sx{s},\sx{s})}
\end{equation}
which follows from the definition of the sets $\sx{s}$ in  proposition~\ref{prop55}
we find the recurrence relation  \r{eq:recZgln-2}  with the functions 
$\Phi_{i,\ell}(\bw)$ and $\Psi^{(\ell)}_{\ell,i}(\bv,z)$ 
defined by the formulas 
\r{eq:phi-ell-gln2} and  \r{eq:psi-ell-gln2}
 respectively.

\subsection{Recurrence relations for $Y(\fg\fl_2)$ and  $Y(\fo_{3})$ models}

It was checked in \cite{LP2020} that the action formulas \r{eq:TijB}
are valid also in the case $n=1$. Similarly, it was verified in \cite{LPR2024}
that the recurrence relations \r{eq:recBBgen} and 
correspondingly \r{eq:recCCgen}  coincide with the recurrence 
relations obtained in \cite{LPRS2019} for $Y(\fo_{3})$ models.

On the other hand  due to the isomorphism 
between Yangians $Y(\fo_{3})$ and $Y(\fg\fl_{2})$
the highest coefficients in $\fo_{3}$ invariant models should be related 
to the  highest coefficients in 
$\fg\fl_{2}$ invariant models. To describe this relation we first 
consider the recurrence relations \r{eq:recZgln-2-ell} and 
\r{eq:recZgln-2} in the case $n=2$, $\ell=1$ and provide 
their solutions in term of the Izergin determinant. 

\paragraph{Izergin determinant.} For a set $\bu$ of  cardinality $|\bu|=r$ we introduce the following triangular 
products of $g$ functions 
\begin{equation}\label{g3}
\delta_g(\bu)=\prod_{1\leq a< b\leq r} g(u_b,u_a),\qquad 
\delta'_g(\bu)=\prod_{1\leq a< b\leq r} g(u_a,u_b)\,.
\end{equation}
The Izergin determinant  
is the rational function $K^{(c)}_r(\bv|\bu)$ depending on two 
sets of formal variables $\bv$ and $\bu$ of the same cardinality 
$|\bv|=|\bu|=r$
\begin{equation}\label{g4}
K^{(c)}_r(\bv|\bu)=h(\bv,\bu)\,\delta_g(\bu)\,\delta'_g(\bv)\ 
{\rm det}\left|\frac{g(v_a,u_b)}{h(v_a,u_b)}\right|_{a,b=1,\ldots,r}\,.
\end{equation}
It satisfies different properties described in the book~\cite{Sl-book}. 
Its dependence on the parameter $c$ occurs through the functions 
$g(u,v)$ and $h(u,v)$. 
In particular, the Izergin determinant 
satisfies recurrence relations which can be written 
in the form 
\begin{equation}\label{g5}
\begin{split}
K^{(c)}_r(\bv,z|\bu)&=f(z,\bu)\sum_{\rm part} 
 \frac{f(\bu_{\st},\bu_{\so})}{h(z,\bu_{\so})}\ K^{(c)}_{r-1}(\bv|\bu_{\st})\,,\\
 K^{(c)}_r(\bu|\bv,z)&=f(\bu,z)\sum_{\rm part} 
 \frac{f(\bu_{\so},\bu_{\st})}{h(\bu_{\so},z)}\ K^{(c)}_{r-1}(\bu_{\st}|\bv)
 \end{split}
\end{equation}
for any sets $\bu$ and $\bv$ with cardinalities $|\bu|=|\bv|+1$. 
In \r{g5} the summation goes over the partitions $\bu\vdash
\{\bu_{\so},\bu_{\st}\}$ such that the cardinality $|\bu_{\so}|=1$. 

\paragraph{Highest coefficients for $\fg\fl_2$ invariant models.} The recurrence relations \r{eq:recZgln-2-ell} and 
\r{eq:recZgln-2} for the highest coefficients $Z(\bu|\bv,z)$ and 
$Z(\bv,z|\bu)$ in $\fg\fl_2$ invariant models takes the form 
\begin{equation}\label{g1}
Z(\bu|\bv,z)=\frac{f(z,\bu)}{h(z,\bv)\,h(\bv,z)}
\sum_{\rm part} \frac{f(\bu_{\st},\bu_{\sth})}{h(z,\bu_{\sth})}\
\frac{Z(\bu_{\st}|\bv)}{h(\bu_{\st},\bu_{\sth})\,h(\bu_{\sth},\bu_{\st})}
\end{equation}
and
\begin{equation}\label{g2}
Z(\bv,z|\bu)=\frac{f(\bu,z)}{h(z,\bv)\,h(\bv,z)}
\sum_{\rm part} \frac{f(\bu_{\so},\bu_{\st})}{h(\bu_{\so},z)}\
\frac{Z(\bv|\bu_{\st})}{h(\bu_{\so},\bu_{\st})\,h(\bu_{\st},\bu_{\so})}
\end{equation}
respectively.
Comparing these equalities with the recurrence relations for the 
Izergin determinant \r{g5} we conclude that they can be solved as follows 
\begin{equation}\label{g6}
Z(\bu|\bv)=\frac{K^{(c)}_r(\bv|\bu)}
{h(\bu,\bu)\,h(\bv,\bv)}\,.
\end{equation}
Note, that for the 'old' normalization of the Bethe vectors $\tilde\BB(\bu)$
and $\tilde\CC(\bv)$ the highest coefficient $\tilde{Z}(\bu|\bv)$
in $Y(\fg\fl_2)$ based model 
coincides with the Izergin determinant $K^{(c)}_r(\bv|\bu)$.

\paragraph{Highest coefficients for $\fo_3$ invariant model.}
Let us write explicitly the recurrence relations for the highest coefficients 
in  $Y(\fo_{3})$ models. 
In that case we have $n=1$, $\ell=0$, $\bar u\equiv \bar u^{(0)}$, 
$\bar v\equiv\bar v^{(0)}$ with cardinalities $|\bv|+1=|\bu|=r$. 

The recurrence relation \r{eq:recZ-other}
given by the proposition~\ref{prop52} takes the 
form 
\begin{equation}\label{o1}
Z(\bu|\bv,z)=\sum_{j=-1}^0\sum_{\rm part} 
\Phi_{1,j}(\bu)\,\Psi^{(0)}_{j,1}(\bv,z)\ Z(\bu_{\st}|\bv_{\st})\,,
\end{equation}
where 
\begin{equation}\label{o2}
\Phi_{1,j}(\bu)\,\Psi^{(0)}_{j,1}(\bv,z)=
\Big(\frac{f(z,\bu)\,\ff(\bu_{\st},\bu_{\sth})}{h(z,\bu_{\sth})}
\Big)\cdot
\Big(\frac{g(z_0,\bv_{\so})\,\ff(\bv_{\so},\bv_{\st})}{\ff(z_0,\bv)}\Big)\,.
\end{equation}
Using \r{o2}  the recurrence relation \r{o1} takes the form 
\begin{equation}\label{o3}
\begin{split}
Z(\bu|\bv,z)&=\frac{f(z,\bu)}{\ff(z_0,\bv)}\sum_{\rm part} 
\frac{\ff(\bu_{\st},\bu_{\sth})}{h(z,\bu_{\sth})}\ Z(\bu_{\st}|\bv)+\\
&\quad+ \frac{f(z,\bu)}{\ff(z_0,\bv)}\sum_{\rm part}
g(z_0,\bv_{\so})\,\ff(\bv_{\so},\bv_{\st})\ 
\frac{\ff(\bu_{\st},\bu_{\sth})}{h(z,\bu_{\sth})}\ 
Z(\bu_{\st}|\bv_{\st})\,,
\end{split}
\end{equation}
where in the first line of \r{o3} sum goes 
over partitions of the set $\bu\vdash\{\bu_{\st},\bu_{\sth}\}$ with cardinalities 
 $|\bu_{\sth}|=1$,  and in the second line of \r{o3} sum goes 
over partitions of the sets $\bv\vdash\{\bv_{\so},\bv_{\st}\}$ 
and $\bu\vdash\{\bu_{\st},\bu_{\sth}\}$ with cardinalities
$|\bv_{\so}|=1$, $|\bu_{\sth}|=2$. 

Analogously one can present the recurrence relation \r{eq:recZ-other2}
given in the proposition~\ref{prop53} for the case $n=1$ and $\ell=0$ 
in the form 
\begin{equation}\label{o4}
Z(\bv,z|\bu)=\frac{\ff(\bu,z)}{\ff(z_0,\bv)}\ \sum_{\rm part} 
g(z_1,\bw_{\so})\,\ff(\bw_{\so},\bw_{\st})\
Z(\bv|\bw_{\st})
\end{equation}
and sum in \r{o4} goes according to 
\r{eq:part-Psi-elli} and  \r{eq:part-Phi-iell} which tell that there is no 
partition of the set $\bv$ and the set $\bw=\{\bu,z_0\}=
\{\bw_{\so},\bw_{\st}\}$ \r{w-set} 
is parted with cardinality $|\bw_{\so}|=2$. 

\begin{prop}\label{prop:so3}
Both recurrence relations \r{o3} and \r{o4} are 
satisfied by the following expression for the highest coefficient
of the scalar product in $Y(\fo_3)$ integrable models
\begin{equation}\label{o5}
Z(\bu|\bv)=2^{|\bu|}\ K^{(c/2)}_r(\bv|\bu)\,.
\end{equation}
\end{prop}
\proof  
The explicit proof of this proposition requires use of the summation 
identity for the Izergin's determinant 
which can be found in the book \cite{Sl-book}. \qed

As it was expected from the isomorphism between 
$Y(\fo_3)$ and $Y(\fg\fl_2)$ at $c\to c/2$ the highest coefficients 
in these models are proportional to the Izergin's determinants 
with the parameters $c/2$ and $c$ respectively.

\section{Analytical properties of the highest coefficients\label{sect:anal}}

Another type of recurrence relations can be obtained when considering the limit ${v^{(s)}_j\to u^{(s)}_j}$ for some $j$ and $s$. It provides the pole structure of the highest coefficient (HC) when the Bethe parameters of the Bethe vector and the dual Bethe vector coincide.
\begin{prop}\label{prop:residue}
For  given fixed $\t=0,1,...,n-1$ and $k=1,2,...,r_\t$, in the limit  ${v^{(\t)}_k\to u^{(\t)}_k}$ we have  the equivalence
\begin{equation}\label{eq:residue}
\begin{split}
Z(\bar u|\bar v)\,{\sim}\ & g(u^{(\t)}_k,v^{(\t)}_k)\,A^{(\t)}_k({\bar u}|\bar v)\,
Z(\mathring{\bar u}|\mathring{\bar v}) +reg.\,,\\
A^{(\t)}_k(\bar u|\bar v)=\ &\gamma_\t(u^{(\t)}_k,{\bar u}^{(\t)}_k)\,\gamma_\t({\bar v}^{(\t)}_k, v^{(\t)}_k) \,
\frac{h(\bar u^{(\t+1)},u^{(\t)}_k)\,h(v^{(\t)}_k,\bar v^{(\t-1)})}{g(u^{(\t)}_k,\bar u^{(\t-1)})\,g(\bar v^{(\t+1)},v^{(\t)}_k)}\\
\ =\ &\Omega(u^{(\t)}_k|\mathring{\bar u})\Omega(\mathring{\bar v}|v^{(\t)}_k)\,
,
\end{split}
\end{equation}
where $\mathring{\bar u}=\bar u\setminus \{u^{(\t)}_k\} $, $\mathring{\bar v}= \bar v\setminus\{v^{(\t)}_k\}$ and $reg.$ denotes terms that are regular in the limit ${v^{(\t)}_k\to u^{(\t)}_k}$.
By convention $\bar u^{(-1)}=\bar v^{(-1)}=\bar u^{(n)}=\bar v^{(n)}=\varnothing$.
\end{prop}
\proof
We prove the property through a recursion on $n$, the rank of $\fo_{2n+1}$.

First we note that for $n=1$, the generalized models for $Y(\fo_3)$ are equivalent to those for $Y(\fg\fl_2)$, see \cite{LPRS2019}.  One can deduce the  scalar products  as
\begin{equation}\label{eq:scal-gl2-o3}
S_{\fo_3}(\bar v|\bar u)= 2^{|\bar u|}\,  \Big( h(\bu,\bu)\,h(\bv,\bv)\ 
S_{\fg\fl_2}(\bar v|\bar u)\Big)_{c\to c/2}\,,
\end{equation}
which was shown in the previous section on the level of the HC.
This implies that all the properties, which have already been proved in the $Y(\fg\fl_2)$ context, are valid for $n=1$. In particular, the residue property is already proven for $n=1$.

\begin{rmk}\label{rem-com}
Assuming that $\bar u^{(0)}=\bar v^{(0)}=\varnothing$, we may also compare the property \eqref{eq:residue} with the Proposition 3.1  of \cite{HLPRS2017b}.
Using the notation of the present article, we rewrite the equation (3.17) of \cite{HLPRS2017b} as 
\begin{equation}\label{eq:res.gln-paper}
\tilde{Z}(\bar u|\bar v)\big\vert_{u^{(\t)}_k\to v^{(\t)}_k}=g(v^{(\t)}_k,u^{(\t)}_k)\,
\frac{f({\bar v}^{(\t)}_k, v^{(\t)}_k)f(u^{(\t)}_k,{\bar u}^{(\t)}_k)}{f({\bar v}^{(\t+1)}, v^{(\t)}_k)f(u^{(\t)}_k,{\bar u}^{(\t-1)})}\,\tilde{Z}(\mathring{\bar u}|\mathring{\bar v})+reg.
\end{equation}

To do the comparison, we need to use the formula \r{c3} which describes
the relation between highest coefficients for the scalar product of the 
Bethe vectors with the normalization used in this paper and the same objects 
for the normalization used in \cite{HLPRS2017b}. 
This difference in normalization produces an 
 additional factor for the residue property of the highest coefficients
\begin{equation*}
\begin{split}
\Xi^{(\t)}_k(\bar u, \bar v) =  & -\prod_{s=1}^{n-1}\frac{h(\bar u^{(s)},\bar u^{(s)})}{h(\bar u^{(s+1)},\bar u^{(s)})}\,
\frac{h(\bar v^{(s)},\bar v^{(s)})}{h(\bar v^{(s+1)},\bar v^{(s)})}\,
\frac{h(\mathring{\bar u}^{(s+1)},\mathring{\bar u}^{(s)})}{h(\mathring{\bar u}^{(s)},\mathring{\bar u}^{(s)})}\,
\frac{h(\mathring{\bar v}^{(s+1)},\mathring{\bar u}^{(v)})}{h(\mathring{\bar v}^{(s)},\mathring{\bar v}^{(s)})}\,
\\
 =\ & -\frac{h({\bar u}^{(\t)}_k, u^{(\t)}_k)h(u^{(\t)}_k,{\bar u}^{(\t)}_k)}{h({\bar u}^{(\t+1)}, u^{(\t)}_k)h(u^{(\t)}_k,{\bar u}^{(\t-1)})}\,
\frac{h({\bar v}^{(\t)}_k, v^{(\t)}_k)h(v^{(\t)}_k,{\bar v}^{(\t)}_k)}{h({\bar v}^{(\t+1)}, v^{(\t)}_k)h(v^{(\t)}_k,{\bar v}^{(\t-1)})}\,.
\end{split}
\end{equation*}
Using the expression \eqref{eq:residue} for $A^{(\t)}_k(\bar u| \bar v)$, it is easy to see that the product $\Xi^{(\t)}_k(\bar u, \bar v)A^{(\t)}_k(\bar u| \bar v)$ just reproduces the factor in 
\eqref{eq:res.gln-paper}.
\end{rmk}

Now, suppose that the property is true for $Y(\fo_{2m+1})$ models with $m<n$, and consider the HC for $Y(\fo_{2n+1})$ models. We show that the relation \eqref{eq:residue} is valid for $Y(\fo_{2n+1})$ models through a recurrence on $r_{n-1}=|\bu^{(n-1)}|$.
When $r_{n-1}=0$, then we are back to a highest coefficient of $Y(\fo_{2n-1})$ models, and the relation is true by the induction
 hypothesis on $n$. We suppose now that it is true for $r_{n-1}<r$ and consider a highest coefficient with $|\bu^{(n-1)}|=r$.

We first look at the case $\t<n-1$, with ${v^{(\t)}_k\to u^{(\t)}_k}$. We consider the recurrence relation \eqref{eq:recZ-other} with $\ell=n-1$. 
Obviously, poles corresponding to the limit ${v^{(\t)}_k\to u^{(\t)}_k}$ may appear only in the highest coefficient $Z(\bar u_{\st}|\bar v_{\st})$ when $v^{(\t)}_k\in\bar v^{(\t)}_{\st}$ and $u^{(\t)}_k\in \bar u^{(\t)}_{\st}$. In that case, we can use the induction hypothesis to get  
\begin{equation*}
Z(\bar u_{\st}|\bar v_{\st})\,{\sim}\  g(u^{(\t)}_k,v^{(\t)}_k)\,A^{(\t)}_k({\bar u_{\st}}|\bar v_{\st})\,
Z(\mathring{\bar u}_{\st}|\mathring{\bar v}_{\st}) +reg.\,,
\end{equation*}
where
$\bar v^{(\t)}_{\so}=\mathring{\bar v}^{(\t)}_{\so}$, $\bar v^{(\t)}_{\st}=\{\mathring{\bar v}^{(\t)}_{\st},v^{(\t)}_k\}$,
with  $\mathring{\bar v}=\bar v\setminus \{v^{(\t)}_k\}$ and similarly for $\bar u$. We have also for 
$0\leq \t\leq n-2$
\begin{eqnarray}
&&\label{eq:psi-rond}
\Psi^{(n-1)}_{j,n}(\bar v, z)=\  \gamma_\t(\bar v^{(\t)}_{\so}, v^{(\t)}_{k})\,
\frac{h( v^{(\t)}_{k},\bar v^{(\t-1)}_{\so})}{g(\bar v^{(\t+1)}_{\so}, v^{(\t)}_{k})}\,\frac{\Psi_{j,n}(\mathring{\bar v}, z)}{\big(g(z, v^{(n-2)}_{k})\big)^{\delta_{p,n-2}}}\,,
\\
&&\label{eq:phi-rond}
\Phi_{n,j}(\bar u,z) =\gamma_\t( u^{(\t)}_{k},\bar u^{(\t)}_{\sth})\,
\frac{h(\bar u^{(\t+1)}_{\sth}, u^{(\t)}_{k})}{g( u^{(\t)}_{k},\bar u^{(\t-1)}_{\sth})}\,\Phi_{n,j}(\mathring{\bar u},z)\,,
\end{eqnarray}
with the convention that $\bar v^{(-1)}=\bar u^{(-1)}=\varnothing$.
This leads to the following formula
\begin{equation*}
\begin{split}
 Z(\bar u|\bar v^{(0)}, \dots, \{\bar v^{(n-1)},z\})=\ & 
g(u^{(\t)}_k,v^{(\t)}_k)\,\sum_{j=-n}^{n-1} 
 \sum_{\rm part}B_{\t,k}^{j,n-1}(\bar u,\bar v,z)\,A^{(\t)}_k({\bar u_{\st}}|\bar v_{\st})\\
&\quad\times\, \Phi_{n,j}(\mathring{\bar u},z) \,\Psi^{(n-1)}_{j,n}(\mathring{\bar v}, z)\, 
 Z(\mathring{\bar u}_{\st}|\mathring{\bar v}_{\st})\ +{reg.}\,,
\end{split}
\end{equation*}
where $B_{\t,k}^{j,n-1}(\bar u,\bar v,z)$ is deduced from formulas \eqref{eq:psi-rond} and \eqref{eq:phi-rond},
and $A^{(\t)}_k({\bar u_{\st}}|\bar v_{\st})$ takes the form \eqref{eq:residue} by the induction hypothesis on $r_{n-1}$.

Now, looking at the expression of $B_{\t,k}^{j,n-1}(\bar u,\bar v,z)$ and $A^{(\t)}_k({\bar u_{\st}}|\bar v_{\st})$, it is easy to see that
\begin{equation}\label{eq:BA=A}
B_{\t,k}^{j,n-1}(\bar u,\bar v,z)\,A^{(\t)}_k(\bar u_{\st}|\bar v_{\st})= A^{(\t)}_k(\bar u|\{\bar v,z^{(n-1)}\})\,.
\end{equation}
When doing the calculation, one has to single out the case  $\t=n-2$, because it makes appear $z$, as shown in \eqref{eq:psi-rond}.
Note that the r.h.s. in \eqref{eq:BA=A} does not depend on the partition, nor on $j$. Then, since 
\begin{equation*}
\begin{split}
\sum_{j=-n}^{n-1} 
 \sum_{\rm part}\Phi_{n,j}(\mathring{\bar u},z) \,\Psi^{(n-1)}_{j,n}(\mathring{\bar v}, z)\, 
 Z(\mathring{\bar u}_{\st}|\mathring{\bar v}_{\st})= Z(\mathring{\bar u}|\mathring{\bar v}^{(0)}, \dots, \{\mathring{\bar v}^{(n-1)},z\})
\end{split}
\end{equation*}
we get the expression for $r_{n-1}=r$. This ends the recurrence for $\t<n-1$.

We now look at the case $\t=n-1$. If $|\bu^{(0)}| = |\bv^{(0)}|=0$, then $Z(\bar u|\bar v^{(1)}, \dots, \bar v^{(n-1)})$ corresponds to a $Y(\fg\fl_n)$ model, for
which the residue property holds. The exact comparison with the results (3.17) of the paper \cite{HLPRS2017b} was presented in the remark~\ref{rem-com}. 

For $|\bu^{(0)}| = |\bv^{(0)}|>0$, we consider the recurrence relation \eqref{eq:recZ-other} with $\ell=0$, in the limit ${v^{(\t)}_k\to u^{(\t)}_k}$, with $\t>1$.
By the recurrence hypothesis, we have
 \begin{equation*}
Z(\bar u_{\st}|\bar v_{\st})\,{\sim}\  g(u^{(\t)}_k,v^{(\t)}_k)\,A^{(\t)}_k({\bar u_{\st}}|\bar v_{\st})\,
Z(\mathring{\bar u}_{\st}|\mathring{\bar v}_{\st}) +reg. 
\end{equation*}
One can also compute
\begin{equation*}
\begin{split}
\Psi_{j,1}^{(0)}(\bar v, z)=\  &\gamma_\t(\bar v^{(\t)}_{\so}, v^{(\t)}_{k})\,
\frac{h( v^{(\t)}_{k},\bar v^{(\t-1)}_{\so})}{g(\bar v^{(\t+1)}_{\so}, v^{(\t)}_{k})}\,\Psi_{j,1}^{(0)}(\mathring{\bar v}, z), \\
\Phi_{1,j}(\bar u,z) =\ &\gamma_\t( u^{(\t)}_{k},\bar u^{(\t)}_{\sth})\,
\frac{h(\bar u^{(\t+1)}_{\sth}, u^{(\t)}_{k})}{g( u^{(\t)}_{k},\bar u^{(\t-1)}_{\sth})}\,\Phi_{1,j}(\mathring{\bar u},z)
\end{split}
\end{equation*}
leading to 
\begin{equation*}
B_{\t,k}^{j,0}(\bar u,\bar v,z)=\gamma_\t(\bar v^{(\t)}_{\so}, v^{(\t)}_{k})\,
\frac{h( v^{(\t)}_{k},\bar v^{(\t-1)}_{\so})}{g(\bar v^{(\t+1)}_{\so}, v^{(\t)}_{k})}\,
\gamma_\t( u^{(\t)}_{k},\bar u^{(\t)}_{\sth})\,
\frac{h(\bar u^{(\t+1)}_{\sth}, u^{(\t)}_{k})}{g( u^{(\t)}_{k},\bar u^{(\t-1)}_{\sth})}\,.
\end{equation*}
Once again we get 
\begin{equation}\label{eq:BA=A2}
B_{\t,k}^{j,0}(\bar u,\bar v,z)\,A^{(\t)}_k(\bar u_{\st}|\bar v_{\st})= A^{(\t)}_k(\bar u|\{\bar v,z^{(0)}\})\,.
\end{equation}
It proves the identity \eqref{eq:residue} for $r_{n-1}=r$ and $1<\t\leq n-1$ and concludes the induction proof. 
\qed

We checked that using the recurrence relation \eqref{eq:recZ-other2} instead of \eqref{eq:recZ-other} leads to the same residue formula \eqref{eq:residue}. As in the proof above, for a given color $\t$, one can consider any recurrence relation 
\eqref{eq:recZ-other2} with $\t\neq\ell-1,\ell,\ell-2$ to do the proof.

\subsection{Residue formula for the scalar product \label{sect:resS}}
From the residue for the highest coefficient, we can deduce a similar relation for the scalar product.
We consider $S(\bar u|\bar v)$ in the limit $u^{(\t)}_k\,\to\, v^{(\t)}_k$.
First, we prove that this limit is non-singular. Then, we show that the coefficient of the derivative of $\alpha_{p}(u^{(\t)}_k)$ can be interpreted as a scalar product of a similar type.
Starting from the sum formula \eqref{eq:scal-fin}, the terms in which $u^{(\t)}_k \in \bar u^{(\t)}_{\so}$ and  $v^{(\t)}_k \in \bar v^{(\t)}_{\st}$ or alternatively $u^{(\t)}_k \in \bar u^{(\t)}_{\st}$ and  $v^{(\t)}_k \in \bar v^{(\t)}_{\so}$ have no singularities in the limit $u^{(\t)}_k\,\to\, v^{(\t)}_k$.
The two contributions with  $u^{(\t)}_k$ (and $v^{(\t)}_k$)  in the subsets $\bar u^{(\t)}_{\so}$ (and $\bar v^{(\t)}_{\so}$) or in the subsets $\bar u^{(\t)}_{\st}$ (and $\bar v^{(\t)}_{\st}$) have a spurious singularity that needs to be dealt with.
In the first case, applying the residue formula for the highest coefficient \eqref{eq:residue}, we get:
\begin{equation}\label{eq:S1}
\begin{split}
S_1(\bar v|\bar u)=&\sum_{\rm part} 
g(v^{(\t)}_k,u^{(\t)}_k) \alpha_p(v^{(p)}_k) A^{(p)}_k(\bar v_{\so}|\bar u_{\so})
\Omega(\bar u_{\so}|{\bar u}_{\st}) \Omega({\bar v}_{\st}|\bar v_{\so})\\
&\times
Z(\mathring{\bar v}_{\so}|\mathring{\bar u}_{\so}) {Z}(\bar u_{\st}|\bar v_{\st})\prod_{s=0}^{n-1}\alpha_s(\mathring{\bar v}^{(s)}_{\so})\alpha_s(\bar u^{(s)}_{\st})+reg.\,,
\end{split}
\end{equation}
where $\mathring{\bar u}={\bar u}\setminus\{u^{(\t)}_{k}\}$ and 
$\mathring{\bar v}={\bar v}\setminus\{v^{(\t)}_{k}\}$.
In the second case we get
\begin{equation}\label{}
\begin{split}
S_2(\bar v|\bar u)=&\sum_{\rm part} 
g(u^{(\t)}_k,v^{(\t)}_k) \alpha_s(u^{(p)}_k) A^{(\t)}_k(\bar u_{\st}|\bar v_{\st})\Omega({\bar u}_{\so}|{\bar u}_{\st}) \Omega({\bar v}_{\st}|{\bar v}_{\so})\\
&\times 
Z(\bar v_{\so}|{\bar u}_{\so}){Z}(\mathring{\bar u}_{\st}|\mathring{\bar v}_{\st})\prod_{s=0}^{n-1}\alpha_s({\bar v}^{(s)}_{\so})\alpha_s(\mathring{\bar u}^{(s)}_{\st})+reg.
\end{split}
\end{equation}
Using the expression \eqref{eq:residue} of $A_k^{(\t)}$, we get
\begin{equation}
\begin{split}
S_1(\bar v|\bar u)&= g(v^{(\t)}_k,u^{(\t)}_k) \alpha_p(v^{(p)}_k) \sum_{\rm part} 
\Omega(v^{(p)}_{k}|\mathring{\bar v}_{\so})
\Omega({\bar v}_{\st}|u^{(p)}_{k})
\Omega(\mathring{\bar u}_{\so}|u^{(p)}_{k})
\Omega(u^{(p)}_{k}|{\bar u}_{\st}) 
\\
&\quad\times
\Omega({\bar v}_{\st}|\mathring{\bar v}_{\so})
\Omega(\mathring{\bar u}_{\so}|{\bar u}_{\st})
Z(\mathring{\bar v}_{\so}|\mathring{\bar u}_{\so}) {Z}(\bar u_{\st}|\bar v_{\st})\prod_{s=0}^{n-1}\alpha_s(\mathring{\bar v}^{(s)}_{\so})\alpha_s(\bar u^{(s)}_{\st})+reg.\,,
\\
S_2(\bar v|\bar u)&= g(u^{(\t)}_k,v^{(\t)}_k) \alpha_s(u^{(p)}_k) \sum_{\rm part} 
 \Omega(v^{(p)}_k|{\bar v}_{\so})  
 \Omega(\mathring{\bar v}_{\st}|v^{(\t)}_k)
 \Omega({\bar u}_{\so}| u^{(p)}_{k})
 \Omega(u^{(\t)}_k|\mathring{\bar u}_{\st})
 \\
&\quad\times 
\Omega(\mathring{\bar v}_{\st}|{\bar v}_{\so})
\Omega({\bar u}_{\so}|\mathring{\bar u}_{\st})
Z(\bar v_{\so}|{\bar u}_{\so}){Z}(\mathring{\bar u}_{\st}|\mathring{\bar v}_{\st})\prod_{s=0}^{n-1}\alpha_s({\bar v}^{(s)}_{\so})\alpha_s(\mathring{\bar u}^{(s)}_{\st})+reg.
\end{split}
\end{equation}
Gathering the two contributions, and using the fact that ${\bar u}_{\st}=\mathring{\bar u}_{\st}$ in $S_1$ and 
${\bar u}_{\so}=\mathring{\bar u}_{\so}$ in $S_2$ (and similarly for $\bar v$), we get
\begin{equation}\label{eq:res_sum0}
\begin{split}
S(\bar v|\bar u)&=\Big(\alpha_\t(v^{(\t)}_{k})- \alpha_\t( u^{(\t)}_{k})\Big)\,g(u^{(\t)}_k,v^{(\t)}_k)\times\\
&\quad\times \sum_{\rm part} 
\Omega(v^{(p)}_k|\mathring{\bar v}_{\so})  
\Omega(\mathring{\bar v}_{\st}|v^{(\t)}_k)
\Omega(\mathring{\bar u}_{\so}| u^{(p)}_{k})
\Omega(u^{(\t)}_k|\mathring{\bar u}_{\st})
\\
&\qquad\times 
\Omega(\mathring{\bar v}_{\st}|\mathring{\bar v}_{\so})
\Omega(\mathring{\bar u}_{\so}|\mathring{\bar u}_{\st})
Z(\mathring{\bar v}_{\so}|\mathring{\bar u}_{\so})
Z(\mathring{\bar u}_{\st}|\mathring{\bar v}_{\st})
\prod_{s=0}^{n-1}\alpha_s(\mathring{\bar v}^{(s)}_{\so})\alpha_s(\mathring{\bar u}^{(s)}_{\st})+reg.
\end{split}
\end{equation}

Expression \eqref{eq:res_sum0} is well-defined in the limit $v^{(\t)}_k\to u^{(\t)}_k$ and we obtain
\begin{equation}\label{eq:res_sum1}
\begin{split}
\lim_{v^{(\t)}_k\to u^{(\t)}_k} S(\bar v|\bar u)=\ &
 -c\, \alpha'_p(u^{(p)}_k)  \sum_{\rm part} 
\Omega(u^{(p)}_k|\mathring{\bar v}_{\so})  
\Omega(\mathring{\bar v}_{\st}| u^{(\t)}_k)
\Omega(\mathring{\bar u}_{\so}| u^{(p)}_{k})
\Omega(u^{(\t)}_k|\mathring{\bar u}_{\st})
\\
&\times 
\Omega(\mathring{\bar v}_{\st}|\mathring{\bar v}_{\so})
\Omega(\mathring{\bar u}_{\so}|\mathring{\bar u}_{\st})
Z(\mathring{\bar v}_{\so}|\mathring{\bar u}_{\so})
Z(\mathring{\bar u}_{\st}|\mathring{\bar v}_{\st})
\prod_{s=0}^{n-1}\alpha_s(\mathring{\bar v}^{(s)}_{\so})\alpha_s(\mathring{\bar u}^{(s)}_{\st})+reg.,
\end{split}
\end{equation} 
where $\mathit{reg}$ means the regular terms do not depend on the derivative of $\alpha_\t$
when $v^{(\t)}_k\to u^{(\t)}_k$.
 
The equation \r{eq:res_sum1} can be considered as 
\begin{equation}\label{eq:res_sum2}
\lim_{v^{(\t)}_k\to u^{(\t)}_k} S(\bar v|\bar u)=  -c\, \alpha'_p(u^{(p)}_k)\, \Omega(\mathring{\bar u}| u_{k}^{(\t)})\Omega(\mathring{\bar v}|u_{k}^{(\t)})\, S^{\rm (mod)}(\mathring{\bar v}|\mathring{\bar u}) + reg.
\end{equation}
where $S^{\rm (mod)}(\mathring{\bar v}|\mathring{\bar u})$ is a scalar product of the form\r{eq:scal-fin} but with other functions:
\begin{equation}\label{alpha:mod}
\alpha^{\rm (mod)}_{s}(z) = \frac{\Omega(u^{(p)}_k|z^{(s)})}{\Omega(z^{(s)}|u^{(p)}_k)} \alpha_{s}(z)\,. 
\end{equation}
More explicitly,
\begin{equation}
\begin{split}
\alpha^{\rm (mod)}_{s}(z)=\alpha_{s}(z)\ \text{ for }\ |s-\t|>1\,,\qquad&
\alpha^{\rm (mod)}_{\t-1}(z) = \alpha_{\t-1}(z)\,f(z, u^{(\t)})\,, \\
\alpha^{\rm (mod)}_{\t}(z)=\alpha_{\t}(z)\,\frac{\gamma_\t(u^{(\t)}_k, z)}{\gamma_\t(z, u^{(\t)}_k)},\qquad&
\alpha^{\rm (mod)}_{\t+1}(z)=
\alpha_{\t+1}(z)\, \frac{1}{f(u^{(\t)}_k, z)}\,.
\end{split}
\end{equation}

Moreover, if we require that $u^{(\t)}_k$ satisfies Bethe equation \eqref{eq:BAE}
\begin{equation}
    \alpha_p(u^{(p)}_k)  \Omega(\mathring{\bar u}| u_{k}^{(\t)}) = \Omega(u_{k}^{(\t)}| \mathring{\bar u})
\end{equation}
we obtain equality
\begin{equation}\label{eq: scpr res}
\lim_{v^{(\t)}_k\to u^{(\t)}_k} S(\bar v|\bar u)=  -c\, \frac{\alpha'_p(u^{(p)}_k)}{\alpha_p(u^{(p)}_k)}\, \, \Omega(u_{k}^{(\t)}| \mathring{\bar u})\, \Omega(\mathring{\bar v}|u_{k}^{(\t)})\, \, S^{\rm (mod)}(\mathring{\bar v}|\mathring{\bar u}) + reg.,
\end{equation}
where $\mathit{reg}$ means the regular terms does not depend on the derivative of $\alpha_\t$ when $v^{(\t)}_k\to u^{(\t)}_k$.

\section{Gaudin formula for the norm\label{sect:gaudin}}

The Gaudin formula expresses the norm of on-shell BVs as a determinant of some matrix, the Gaudin matrix \cite{Gaudin}.
This property has been proven in the context of $Y(\fg\fl_2)$ models in \cite{Kore} and in \cite{Resh1986} for $Y(\fg\fl_3)$ models.
For $Y(\fg\fl_n)$ and $Y(\fg\fl_{m|n})$ models, see \cite{HLPRS2017b} for a general presentation. This implies that for $Y(\fo_{2n+1})$ models, the property is proven for BVs of the form $\BB(\varnothing,\bar u^{(1)},\dots,\bar u^{(n-1)})$, since they correspond to $Y(\fg\fl_n)$ BVs.
Moreover, since  generalized models for $Y(\fo_3$) are equivalent to those for $Y(\fg\fl_2)$, see \cite{LPRS2019}, it is also valid for $n=1$.
The goal of this section is to prove that the Gaudin formula is valid for all on-shell BVs in $Y(\fo_{2n+1})$ models.

\subsection{Gaudin matrix for orthogonal models}
We start with the Bethe ansatz equations (BAEs):
\beano
\alpha_s(\bar u^{(s)}_{\so}) &=& \frac{f(\bar u^{(s)}_{\so},\bar u^{(s)}_{\st})}{f(\bar u^{(s)}_{\st},\bar u^{(s)}_{\so})}
\frac{f(\bar u^{(s+1)},\bar u^{(s)}_{\so})}{f(\bar u^{(s)}_{\so},\bar u^{(s-1)})},
\quad s=1,\ldots,n-1,\\
\alpha_0(\bar u^{(0)}_{\so}) &=& \frac{\ff(\bar u^{(0)}_{\so},\bar u^{(0)}_{\st})}{\ff(\bar u^{(0)}_{\st},\bar u^{(0)}_{\so})}
f(\bar u^{(1)},\bar u^{(0)}_{\so}),
\eeano
that hold for arbitrary partitions of the sets $\bar u^{(s)}$ into subsets
$\{\bar u^{(s)}_{\so},\;\bar u^{(s)}_{\st}\}$,
with $\bar u^{(-1)}=\varnothing=\bar u^{(n)}$.

We introduce
\begin{equation*}
\begin{split}
\Gamma^{(s)}_j(\bar u)&=\alpha_s( u^{(s)}_{j}) \, \frac{f(\bar u^{(s)}_{j}, u^{(s)}_{j})}{f( u^{(s)}_{j},\bar u^{(s)}_{j})}
\frac{f( u^{(s)}_{j},\bar u^{(s-1)})}{f(\bar u^{(s+1)}, u^{(s)}_{j})},
\quad j=1,...,r_s\,,\  s=1,...,n-1,\\
\Gamma^{(0)}_j(\bar u) &=\alpha_0( u^{(0)}_{j})\, \frac{\ff(\bar u^{(0)}_{j}, u^{(0)}_{j})}{\ff( u^{(0)}_{j},\bar u^{(0)}_{j})f(\bar u^{(1)}, u^{(0)}_{j})}
\,,\qquad j=1,...,r_0\,,
\end{split}
\end{equation*}
so that the BAEs for the partitions $\bar u^{(s)}\vdash \{u_j^{(s)},\bar u^{(s)}_j\}$ read $\Gamma^{(s)}_j(\bar u)=1$.

The Gaudin matrix $G(\bar u)$ for the $\fo_{2n+1}$ models is a block matrix 
\begin{equation*}
    G(\bar u)=\big(G^{(s,\t)}(\bar u)\big)_{s,\t=0,1,...,n-1}\,,
\end{equation*} 
where the block 
$G^{(s,\t)}(\bar u)$ has size $r_s\times r_\t$. The entries in each block are defined by
\begin{equation}\label{def:G}
G^{(s,\t)}_{j,k}(\bar u) = -c \frac{d}{du^{(\t)}_k} \ln\Gamma^{(s)}_j(\bar u)\,.
\end{equation}
Let us note that in the context of generalized models the functions $\alpha_s$ are free, so that $\Gamma^{(s)}_j$ is a function depending on two sets of variables $\bar u$ and $\bar \alpha=\{\alpha_s(u_j^{(s)}),\ 1\leq j\leq r_s\,,\ 0\leq s\leq n-1\}$, considered as independent variables, so that we should write $\Gamma^{(s)}_j(\bar u;\bar \alpha)$ instead of $\Gamma^{(s)}_j(\bar u)$. We keep the latter notation to lighten the presentation. In the same way, the matrix $G$ depends on two sets of variables $\bar u$ and $\bar X$ with 
\begin{equation*}
X^{(s)}_j=-c\,\frac{d}{dz}\, \ln\alpha_s(z)\Big|_{z=u^{(s)}_j}\,,\quad 1\leq j\leq r_s\,,\ 0\leq s\leq n-1\,.
\end{equation*}
To present the explicit expression of the entries of $G$, we introduce the rational functions 
\begin{equation*}
K^{(s)}(x,y)= \frac{2c_sc}{(x-y)^2-c_s^2}\,,\quad I(x,y)=\,\frac{c^2}{(x-y+c)(x-y)}\,,\quad c_s=\Big(1-\frac{\delta_{s,0}}{2}\Big)c.
\end{equation*}
Then, we have
\begin{equation}\label{eq:G}
\begin{split}
G^{(s,p)}_{j,k}(\bar u) = &\, 0,\quad\text{ when}\quad |s-p|>1\,,
\\
G^{(s,s-1)}_{j,k}(\bar u) =&-I(u^{(s)}_j,u^{(s-1)}_k)\,,
\\
G^{(s,s+1)}_{j,k}(\bar u) =&-I(u^{(s+1)}_k,u^{(s)}_j)\,,
\\
G^{(s,s)}_{j,k}(\bar u) =&\, \delta_{jk}\left(
X^{(s)}_j-\sum_{\ell=1}^{r_s}K^{(s)}(u^{(s)}_j,u^{(s)}_\ell) +\right.\\
&+\left.\sum_{\ell=1}^{r_{s-1}}I(u^{(s)}_j,u^{(s-1)}_\ell)
+\sum_{\ell=1}^{r_{s+1}}I(u^{(s+1)}_\ell,u^{(s)}_j) 
\right) +K^{(s)}(u^{(s)}_j,u^{(s)}_k)\,.
\end{split}
\end{equation}
Notice that for any $s$ and $j$, we have the relation
\begin{equation}\label{eq:X=sumG}
\sum_{p=0}^{n-1} \sum_{k=1}^{r_p}G^{(s,p)}_{j,k}(\bar u) = X^{(s)}_j\,.
\end{equation}

Due to the property of BVs, and the results obtained in the $\fg\fl_n$ models, we know that 
$G(\varnothing,\bar u^{(1)},...,\bar u^{(n-1)})=G_{\fg\fl_n}(\bar u^{(1)},...,\bar u^{(n-1)})$, where $G_{\fg\fl_n}(\bar u)$ is the Gaudin matrix for $\fg\fl_n$ models, see \cite{Kore,MV,HLPRS2017b}.

\subsection{Korepin criteria for orthogonal models}
The Korepin criteria is a list of property for a series of functions $F^{\bar r}$ of  $2\Vert\bar r\Vert$ variables, with $\bar r=(r_0,r_1,\ldots,r_{n-1})$ and 
$\Vert\bar r\Vert=r_0+r_1+\ldots+r_{n-1} >0$:
\begin{defi}[Korepin criteria]\label{def:Kore}
Let $F^{\bar r}$, be a series of functions of $2\Vert\bar r\Vert$ variables $(\bar X;\bar u)$. We say that the functions $F^{\bar r}$ obey the Korepin criteria if they obey the following properties
\begin{enumerate}
\item The functions $F^{\bar r}$ are symmetric under the exchange  of any pairs $(X^{(s)}_j,u^{(s)}_j)\, \leftrightarrow\, (X^{(s)}_k,u^{(s)}_k)$.
\item  The functions $F^{\bar r}$ are polynomials of degree 1 in each $X^{(s)}_j$.
\item For $\Vert\bar r\Vert=1$ with $s$ the unique color such that $r_s=1$, we have $F^{\bar r}(X^{(s)}_1,u^{(s)}_1) = X^{(s)}_1$.
\item The coefficient of $X^{(s)}_j$ in $F^{\bar r}(\bar X;\bar u)$ is equal to a function $F^{\bar r'}(\bar Y;\bar v)$:
\begin{equation}
\frac{\partial\,F^{\bar r}(\bar X;\bar u)}{\partial\,X^{(s)}_j} = F^{\bar r'}(\bar Y;\bar v)\,,
\end{equation}
where the variables entering the new function $F^{\bar r'}(\bar Y;\bar v)$ are expressed as
\begin{equation}\begin{split}
&\bar r'=(r_0,r_1,..., r_{s-1},r_s-1,r_{s+1},..., r_{n-1}),\\
&v^{(s)}_k=u^{(s)}_k \quad \text{;}\quad Y^{(s)}_k=X^{(s)}_k-K^{(s)}(u^{(s)}_j,u^{(s)}_{k})\,,\quad 1\leq k<j \,,\\
&v^{(s)}_k=u^{(s)}_{k+1} \quad \text{;}\quad Y^{(s)}_k=X^{(s)}_{k+1}-K^{(s)}(u^{(s)}_j,u^{(s)}_{k+1})\,,\quad 
j\leq k< r_s\,,\\
&v^{(s+1)}_k=u^{(s+1)}_k \quad \text{;}\quad
Y^{(s+1)}_k=X^{(s+1)}_k+I(u^{(s+1)}_k,u^{(s)}_j)\,,\quad 1\leq k\leq r_{s+1}\,,\\
&v^{(s-1)}_k=u^{(s-1)}_k\quad \text{;}\quad
Y^{(s-1)}_k=X^{(s-1)}_k + I(u^{(s)}_j,u^{(s-1)}_k)\,,\quad 1\leq k\leq r_{s-1}\,,\\
&v^{(\t)}_k=u^{(\t)}_{k}\quad \text{;}\quad Y^{(\t)}_k=X^{(\t)}_k \,,\quad 1\leq k\leq r_\t
\,,\qquad |s-\t]>1\,.
\end{split}\end{equation}
\item $F^{\bar r}(\bar X;\bar u)=0$ if all $X^{(s)}_j=0$.
\end{enumerate} 
\end{defi} 
The Korepin criteria allows a characterization of the functions $F^{\bar r}$:
\begin{lemma}\label{lem:Kore}
The Korepin criteria fixes uniquely the functions $F^{\bar r}$. 
\end{lemma}
\proof
Let $F^{\bar r}$ and $\tilde F^{\bar r}$ be two series of functions obeying the Korepin criteria. We prove by recursion on $\Vert\bar r\Vert$ that the functions $F^{\bar r}$ and $\tilde F^{\bar r}$ coincide. 

By point 3. we have immediately that for $\Vert\bar r\Vert=1$, $F^{\bar r}=\tilde F^{\bar r}$. Suppose now that when $\Vert\bar r\Vert\leq r$, we have 
$F^{\bar r}=\tilde F^{\bar r}$ and consider ${\bar r}$ with $\Vert\bar r\Vert=r+1$, $F^{\bar r}$ and $\tilde F^{\bar r}$. 

By point 4. and the recursion hypothesis, we deduce that
\begin{equation*}
\frac{\partial}{\partial\,X^{(s)}_j}\Big(F^{\bar r}(\bar X;\bar u)-\tilde F^{\bar r}(\bar X;\bar u)\Big)= 0\,,
\end{equation*}
Since by point 2. the functions are of degree 1 in $X^{(s)}_j$, it implies that they coincide up to a term independent from $\bar X$. 
Finally, point 5. shows that they coincide exactly at $\bar X=0$, which ends the proof.
\qed

For our purpose, we define the functions
\begin{equation}\label{def:F}
F^{\bar r}(\bar X;\bar u) = \text{det}\, G(\bar u)\,,\quad \Vert\bar r\Vert=|\bar u|=|\bar X|\,.
\end{equation}
For generalized models, where no specific representations have been chosen, the variables $\bar X$ are independent from the variables $\bar u$, so that $F^{\bar r}$ is a true function of $2\Vert\bar r\Vert$ variables. 
\begin{prop}\label{prop:detG}
The functions $F^{\bar r}$ defined in \eqref{def:F} satisfy the Korepin criteria. 
\end{prop}
\proof
We prove the properties of the criteria point by point.

Point 1: The exchange of any pair amounts to exchange the two corresponding lines and the two corresponding column in the matrix $G$. Then, since each $F^{\bar r}$ is a determinant, it is invariant under this exchange.

Points 2. and 3. are obvious from the explicit form \eqref{eq:G} of $G$. 

To get the point 4, when computing $\frac{\partial\,F^{\bar r}(\bar X;\bar u)}{\partial\,X^{(s)}_j}$, we expand the determinant along the column containing $X^{(s)}_j$, showing that $\frac{\partial\,F^{\bar r}(\bar X;\bar u)}{\partial\,X^{(s)}_j}$ is just the minor of the diagonal element containing $X^{(s)}_j$. It is thus a determinant of the same form, but with a modification of the variables $X^{(p)}_k$ and $u^{(p)}_{k}$ as mentioned. The shift of the variables $Y^{(p)}_k$ allows to add the terms $K^{(s)}(u^{(s)}_j,u^{(s)}_{k})$, $I(u^{(s+1)}_k,u^{(s)}_j)$ and $I(u^{(s)}_j,u^{(s-1)}_k)$ that should appear in the sums occurring in the diagonal terms.

Point 5. follows from the relation \eqref{eq:X=sumG}, which implies the vanishing of $\mathrm{det}\, G$.
\qed

\subsection{Norm of on-shell BVs for orthogonal models}
\begin{thm}\label{thm:gaud}
Let $\BB(\bar u)$ be an on-shell Bethe vector, and $\CC(\bar v)$ a dual Bethe vector with $|\bar v|=|\bar u|$. Then,
\begin{equation} 
\begin{split}
\lim_{\bar v\to\bar u} S(\bar v | \bar u) &=
\lim_{\bar v\to\bar u}\CC(\bar v)\BB(\bar u) = 
\prod_{s,p=0}^{n-1} \prod_{k=1}^{r_s} \prod_{l=1}^{r_p} \Omega(u^{(s)}_k|u^{(p)}_l)\, 
{\rm det}\,G(\bar u)\,
\\
&= \prod_{s=0}^{n-1} \prod_{k,l=1\atop k\ne l}^{r_s} \gamma_s(u^{(s)}_k,u^{(s)}_l) \prod_{s=1}^{n-1} \frac{h(\bu^{(s)},\bu^{(s-1)})}{g(\bu^{(s)},\bu^{(s-1)})} \, {\rm det}\, \, G(\bar u)\,,
\end{split}
\end{equation}
where $G(\bar u)$ is the Gaudin matrix \eqref{def:G}.
\end{thm}
\proof
Let
\begin{equation}\label{eq:Ndef}
 \mathcal{N}(\bu) = 
\prod_{s,p=0}^{n-1} \prod_{k=1}^{r_s} \prod_{l=1}^{r_p} \Omega(u^{(s)}_k|u^{(p)}_l)^{-1} \
\lim_{\bar v\to\bar u}\CC(\bar v)\BB(\bar u),  
\end{equation} 
where $\bu$ satisfy Bethe equation \eqref{eq:BAE}.

From lemma \ref{lem:Kore} and property \ref{prop:detG}, it is sufficient to prove that the $\mathcal{N}(\bu)$ obeys the 
Korepin criteria in the limit $\bar v\to\bar u$. 

\begin{itemize}
\item{Point 1.} It is a direct consequence of the symmetry property for the Bethe vectors.
\item{Point 2.} It is the result of the  equation \eqref{eq: scpr res}.
\item{Point 3.} We start with the Bethe vectors and dual Bethe vectors
\begin{equation*}
\BB(u^{(s)}_1)=\frac{T_{s,s+1}(u^{(s)}_1)}{\lambda_{s+1}(u^{(s)}_1)}|0\rangle
\quad\text{and}\quad \CC(v^{(s)}_1)=\langle0|\,\frac{T_{s+1,s}(v^{(s)}_1)}{\lambda_{s+1}(v^{(s)}_1)},
\end{equation*}
where for $\bar u=\{\varnothing,\dots,\varnothing, u^{(s)}_1,\varnothing,\dots,\varnothing\}$, we wrote $\BB(u^{(s)}_1)$ for $\BB(\bar u)$ and similarly for $\CC(v^{(s)}_1)$.
From the commutation relations \eqref{comtt} and the conditions \eqref{eq:hw}, it is easy to get
\begin{equation*}
\CC(v^{(s)}_1)\BB(u^{(s)}_1) = -\ c\,\frac{\alpha_s(u^{(s)}_1)-\alpha_s(v^{(s)}_1)}{u^{(s)}_1-v^{(s)}_1}
\end{equation*}
which leads to
\begin{equation*} 
\lim_{v^{(s)}_1\to u^{(s)}_1}\CC(v^{(s)}_1)\BB(u^{(s)}_1) = -\ c\,\frac{d}{dz}\alpha_s(z)\Big|_{z=u^{(s)}_1}=\alpha_s(u^{(s)}_1)X^{(s)}_1\,.
\end{equation*}
Now, the Bethe equations \eqref{eq:BAE} in this case read $\alpha_s(u^{(s)}_1)=1$, so that the scalar product is exactly $X_1^{(s)}$ in the limit $v^{(s)}_1\to u^{(s)}_1$.
Moreover, since there is only one Bethe parameter, the normalisation factor in \eqref{eq:Ndef} is 1, so that we get the result.
\item{Point 4.} It follows directly from formula \eqref{eq: scpr res}, taking into account the normalization in \eqref{eq:Ndef}.
\item{Point 5.} When $X^{(s)}_j=0$, $\alpha_s(v^{(s)}_j) = \alpha_s(u^{(s)}_j) + O\Big((v^{(s)}_j-u^{(s)}_j)^2\Big)$, we can replace $\alpha_s(v^{(s)}_j)$ by $\alpha_s(u^{(s)}_j)$, so that the sum formula \eqref{eq:scal-fin} rewrites
\begin{equation}\label{eq:scal4}
\begin{split}
S(\bar v|\bar u)=&\left(\prod_{s=0}^{n-1}\alpha_s(\bar u^{(s)})\right) \lim_{\bar v\to \bar u}S_0(\bar v|\bar u)\,,
\\
S_0(\bar v|\bar u) =&\sum_{\rm part} Z(\bar v_{\so}|\bar u_{\so})\, {Z}(\bar u_{\st}|\bar v_{\st}) \\
&\times
\prod_{s=0}^{n-1} \gamma_s(\bar u^{(s)}_{\so},\bar u^{(s)}_{\st})\,\gamma_s(\bar v^{(s)}_{\st},\bar v^{(s)}_{\so})
\prod_{s=0}^{n-1}\frac{ h(\bar u^{(s+1)}_{\st},\bar u^{(s)}_{\so})h(\bar v^{(s+1)}_{\so},\bar v^{(s)}_{\st})}{g(\bar u^{(s+1)}_{\so},\bar u^{(s)}_{\st})g(\bar v^{(s+1)}_{\st},\bar v^{(s)}_{\so})}\,.
\end{split}
\end{equation}
$S_0(\bar v|\bar u)$ is a particular case of scalar product where all $\alpha_s(z)=1$. This corresponds to a particular model where $T_{i,j}(z)=\delta_{ij}$. But in that case, the only non zero Bethe vector is $|0\rangle$, which is not considered here\footnote{We recall that the Korepin criteria applies to functions $F^{\bar r}$ with $\Vert\bar r\Vert>0$.}. This implies that $S_0(\bar v|\bar u) =0$ as soon as $|\bar u|>0$, and proves Point 5.\qed
\end{itemize}

\section{Conclusion\label{conclu}}
Now that scalar products and norms of Bethe vectors for $\mathfrak{o}_{2n+1}$ models have been computed, several promising directions for future research can be identified. Naturally, the extension to $\mathfrak{o}_{2n}$ and $\mathfrak{sp}_{2n}$ models is a necessary next step. Surprisingly, preliminary investigations suggest that while the algebraic structure remains comparable, the technical aspects of the computations become significantly more intricate. At present, we do not possess a clear interpretation of this discrepancy, but we aim to revisit this issue in future work.

Furthermore, the deformed case -- namely, quantum affine algebras associated with orthogonal and symplectic Lie algebras -- should also be investigated. Similarly, the superalgebra versions of these models, involving $\mathfrak{osp}_{m|2n}$-invariant models, are also worth studying. In particular, the case of $\mathfrak{osp}_{1|2}$-invariant models is of special interest since it is the 'supersymmetric' version of the celebrated XXX ($\mathfrak{gl}_{2}$-invariant) model. We expect that most of the properties of XXX model (determinant form for scalar products and form factors, see below) to be still valid for this 'supersymmetric XXX model'.

Another important research avenue is the computation of form factors, and ultimately, correlation functions. The techniques developed for $\mathfrak{gl}_n$-invariant models are applicable in this context. First, one can compute diagonal form factors -- i.e., those corresponding to diagonal entries of the monodromy matrix -- by employing a twisted transfer matrix; see~\cite{PRS2015a} for results in the $\mathfrak{gl}_3$ case. Subsequently, off-diagonal form factors may be obtained from diagonal ones via the zero mode method, as detailed in~\cite{PRS2014a}.

Within this program, the thermodynamic limit $L \to \infty$ is of fundamental importance for the evaluation of correlation functions. A determinant representation of scalar products of Bethe vectors becomes essential for taking this limit. Once such determinant expressions are obtained, they directly facilitate the computation of form factors, as outlined above. Unfortunately, explicit determinant formulas are known only in limited cases: for instance, $\mathfrak{gl}_2$ models and their quantum deformations (such as the XXZ spin chain) \cite{Slavnov89}, for $\mathfrak{gl}_3$ models \cite{BPRS12} and their quantum deformation \cite{Slavnov2015}, 
and for $\mathfrak{gl}_{2|1}$ models \cite{HLPRS16}. More general expressions are still lacking. However, one may hope that existing formulations of scalar products as sums over partitions could be transformed into determinants of block matrices, in the same way it has been done in \cite{BPRS12,Slavnov2015}.

Detailed studies of specific models are also highly desirable. For instance, a Bose gas model based on $\mathfrak{so}_3$ symmetry -- essentially a spin-1 extension of the ``standard'' Bose gas studied four decades ago~\cite{IzKoRe87} -- would be of considerable interest. 
Notice that the correspondence established in proposition \ref{prop:so3} of the present paper ensures that the scalar products of $Y(\fo_3)$ models have a determinant form. Comparison between this $\mathfrak{o}_3$ bose gas, and the two component bose gase based on $\mathfrak{gl}_3$ and studied in \cite{PRS2015b} would be very interesting. Additionally, the $\mathfrak{o}_5$ model is particularly noteworthy, as it is the first instance in which a genuine departure from the structure of $\mathfrak{gl}_n$ models occurs. 
Since $\mathfrak{o}_5$ is a rank 2 algebra, like $\mathfrak{gl}_3$, one may hope to carry analytical computations as far as has been achieved for $\mathfrak{gl}_3$. Unique features distinguishing the various symmetry classes should begin to emerge in such a study.

\section*{Acknowledgement}
S.P. acknowledges support from the PAUSE Programme and the hospitality of LAPTh, where this work was conducted.
A.L. acknowledges support from the Beijing Natural Science Foundation (Grant No. IS24006). 
He also thanks CNRS PHYSIQUE for support during his visit to LAPTh in the course of this research.

\appendix

\section{Proof of the lemma~\ref{lem42}}\label{ApA}
We prove this lemma  using a
recursion on $n$, the rank of $\fo_{2n+1}$. Since the property has been already proved for $\fg\fl_2$, it is also true for $\fo_3$, i.e. when $n=1$.

Through iterative applications of the relation \eqref{eq:recCCgen} with $\ell=n-1$, we can express a dual BV as
\begin{equation}\label{eq:recCC-multi}
\begin{split}
\CC(\bar v^{(0)}, \dots, \bar v^{(n-1)})&=
\sum_{j_1,\ldots,j_{r_{n-1}}}\sum_{\rm part}
\Psi^{(n-1)}_{j_1\dots j_{r_{n-1}}}(\bar v^{(0)}, \dots, \bar v^{(n-1)})\,\\
&\qquad\times\,\CC(\bar v^{(0)}_{\st},  \dots, \bar v^{(n-2)}_{\st})\, 
\frac{T_{n,j_{r_n}}(v^{(n-1)}_{r_{n-1}})\cdots T_{n,j_1}(v^{(n-1)}_{1})}{\lambda_n(\bar v^{(n-1)})}\,,
\end{split}
\end{equation}
where $\Psi^{(n-1)}_{j_1\dots j_{r_{n-1}}}(\bar v^{(0)}, \dots, \bar v^{(n-1)})$ are rational functions that do not depend on the eigenvalues $\lambda_j(v)$, $ j=-n,\dots,n$, where  $r_{n-1}=\big|\bar v^{(n-1)}\big|$. In \eqref{eq:recCC-multi}, the sum 
runs on all values of $j_1,\dots ,j_{r_{n-1}}=-n,\dots,n-1$ and over partitions of the sets
$\bar v^{(s)}\vdash \{\bar v^{(s)}_{\so},\bar v^{(s)}_{\st}\}$, $s=0,\dots,n-1$.

In the same way, from the action formula \eqref{eq:TijB}, we get by iteration
\begin{equation}\label{eq:multBB}
\frac{T_{n,j_{r_{n-1}}}(v^{(n-1)}_{r_{n-1}})\cdots T_{n,j_1}(v^{(n-1)}_{1})}{\lambda_n(\bar v^{(n-1)})}\,\BB(\bar u) = 
\sum_{\rm part} \left(\prod_{s=0}^{n-1}\alpha_s(\bar w^{(s)}_{\sth})\right)\Phi_{j_1,\dots,j_{r_{n-1}}}(\bar w) \,\BB(\bar w_{\st})\,,
\end{equation}
with now $\bar w^{(s)}=\{\bar u^{(s)},\bar v^{(n-1)},\bar v^{(n-1)}-c(s-1/2)\}$ with $s = 0, \ldots, n-2$. This shows that the functions $\alpha_s$ appearing in the r.h.s. of \eqref{eq:multBB} do not depend on the parameters $\{\bar v^{(0)},\dots,\bar v^{(n-2)}\}$.

Moreover, each operator $T_{n,j}(u)$ contributes to the color $n-1$ by a factor
$$
\begin{cases} -2, &\text{ if }\ j=-n,\\
-1, & \text{ if }\ |j|<n.
\end{cases}
$$
Hence, the r.h.s. in \eqref{eq:multBB} is formed of BVs with no color $n-1$, and as such corresponds to an $\fo_{2n-1}$ model with Bethe parameters 
$\bar w_{\st}\subset\{\bar u^{(s)},\bar v^{(n-1)},\bar v^{(n-1)}-c(s-1/2)\}$. Then, by the recursion hypothesis, we know that the scalar products
$\CC(\bar v^{(0)}_{\st},  \dots, \bar v^{(n-2)}_{\st})\,\BB(\bar w_{\st})$ depend only on $\alpha_s(\bar v^{(s)}_{\st})$ and $\alpha_s(\bar w^{(s)}_{\st})$. This shows that for $s=0, \dots, n-2$, the set $\bar v^{(s)}$ enter only in the functions $\alpha_s$. This property is preserved by the product occurring in the r.h.s. of \eqref{eq:multBB}. By symmetry in the exchange $\bar u\,\leftrightarrow\,\bar v$ this property also applies to the sets $\bar u^{(s)}$, $s=0, \dots, n-2$.

It remains to show that the property is also valid for the sets $\bar u^{(n-1)}$ and $\bar v^{(n-1)}$. For such a purpose, we perform the same calculation as above with  the recursion \eqref{eq:recCCgen} for an integer $\ell$ such that\footnote{Note that if such $\ell$ does not exist, we are dealing with $\mathfrak{gl}_2$ BVs with Bethe parameters $\bar v^{(n-1)}$, for which the lemma \ref{lem42} has been already proven.} $\ell\neq n-1$ and $|\bar v^{(\ell)}|\neq0$.  To simplify the presentation, we take $\ell=n-2$, the other cases being similar. 
\begin{equation}\label{eq:recCC2}
\begin{split}
&\CC(\bar v^{(0)}, \dots, \bar v^{(n-2)}, \bar v^{(n-1)})=
\,
\\
& 
\sum_{\atopn{-n\leq j_1,\ldots,j_{r_{n-2}}\leq n-2}{n-1\leq i_1,\ldots,i_{r_{n-2}}\leq n}}
\sum_{\rm part} \alpha_{n-1}(\bv_{\sth}^{(n-1)})
\Psi^{(n-2)}_{\bar \imath,\bar \jmath}(\bv)\,
\CC(\bar v_{\st}^{(0)},  \dots, \bar v_{\st}^{(n-3)}, \varnothing,\bar v_{\st}^{(n-1)})
\cO_{\bar \imath,\bar \jmath}(\bar v^{(n-2)})\,,
\\
&\cO_{\bar \imath,\bar \jmath}(\bar v^{(n-2)})= \frac{T_{i_1,j_1}(v^{(n-2)}_{1})\cdots T_{i_{r_{n-2}},j_{r_{n-2}}}(v^{(n-2)}_{r_{n-2}})}{\lambda_{n-1}(\bar v^{(n-2)})}
\,,
\end{split}
\end{equation}
where $\bar \imath$ (resp. $\bar \jmath$) stands for $i_1,...,i_{r_{n-2}}$ (resp. $j_1,...,j_{r_{n-2}}$). 

It remains to act with $\cO_{\bar \imath,\bar \jmath}(\bar v^{(n-2)})$ on $\BB(\bar u)$ using iteratively the action formula \eqref{eq:TijB}. As fas as $\alpha_s$ functions are concerned, it produces terms 
$\prod_{s=0}^{n-1}\alpha_s(\bar w^{(s)}_{\so})$, where
$\bar w^{(s)}=\{\bar u^{(s)}, \bar v^{(n-2)}, \bar v^{(n-2)}-c(s-1/2)\}$. This shows that the set $\bar v^{(n-1)}$ enters only in the $\alpha_{n-1}$ function, as expected. By symmetry in the exchange $\bar u\,\leftrightarrow\,\bar v$ this property also true for the set $\bar u^{(n-1)}$.

Finally, from the construction done above, it is clear that each Bethe parameter
$v^{(s)}_i$ and each Bethe parameter $u^{(s)}_j$ occurs at most once in $\alpha^{(s)}$.
\qed

\section{Existence of a suitable model\label{sect:mod}}
We first remind that from Theorem 5.16 in \cite{AMR05}, the  finite dimensional irreducible modules of $Y(\fo_{2n+1})$ are classified by monic polynomials $P_0(u)$, \dots, $P_{n-1}(u)$ such that
\begin{equation}
 \frac{\lambda_{s}(u)}{\lambda_{s+1}(u)}=\frac{P_{s}(u+c)}{P_s(u)}\,,\ s=1, \dots, n-1\mb{and}
 \frac{\lambda_{0}(u)}{\lambda_1(u)}=\frac{P_{0}(u+c/2)}{P_0(u)}.
\end{equation}
We are looking for a module where relations \eqref{eq:alphas} are obeyed. We denote $p_s=|\bar u^{(s)}_{\rm i}|$ and $q_s=|\bar v^{(s)}_{\rm ii}|$.
Let us consider two irreducible modules:
\begin{enumerate}
\item One with polynomials
\begin{equation}\label{eq:polyP}
P_s(z)=\prod_{k=1}^{p_s}(z-c-u^{(s)}_{k})\,,\ s=1,\dots, n\mb{and} P_0(z)=\prod_{k=1}^{p_0}(z-c/2-u^{(0)}_{k}),
\end{equation} 
where $\bar u^{(s)}_{\rm i}=\{u^{(s)}_{k}\,,\ k=1,  \dots, p_s\}$, and associated to the monodromy matrix $T^{(1)}(u)$.
\item One with polynomials
\begin{equation}\label{eq:polyQ}
Q_s(z)=\prod_{k=1}^{q_s}(z-c-v^{(s)}_{k})\,,\ s=1,\dots, n\mb{and} Q_0(z)=\prod_{k=1}^{q_0}(z-c/2-v^{(0)}_{k}),
\end{equation}
where
$\bar v^{(s)}_{\rm ii}=\{v^{(s)}_{k}\,,\ k=1, \dots, q_s\}$, and associated to the monodromy matrix $T^{(2)}(u)$.
\end{enumerate}
We consider the composite model based on $T^{(1)}(u)$ and $T^{(2)}(u)$. By construction, we have 
\begin{equation}
\begin{split}
&\alpha^{[1]}_{s}(u)=\frac{Q_{s}(u+c)}{Q_s(u)}\,,\ s=1, \dots, n-1\mb{and}
 \alpha^{[1]}_{0}(u)=\frac{Q_{0}(u+c/2)}{Q_0(u)},\\
&\alpha^{[2]}_{s}(u)=\frac{P_{s}(u+c)}{P_s(u)}\,,\ s=1, \dots, n-1\mb{and}
 \alpha^{[2]}_{0}(u)=\frac{P_{0}(u+c/2)}{P_0(u)}.\\
\end{split}
\end{equation}
In view of the expressions \eqref{eq:polyP} and \eqref{eq:polyQ}, for generic values of the Bethe parameters, we clearly have 
\begin{equation}
\alpha^{[1]}_s(z)=0\mb{if} z\in\bar v^{(s)}_{\rm ii} \mb{and} \alpha^{[2]}_s(z)=0\mb{if} z\in\bar u^{(s)}_{\rm i}\,,
\end{equation}
as required.

\end{document}